\documentclass{mn2e}
\usepackage{psfig}

\def\ltsima{$\; \buildrel < \over \sim \;$}
\def\lsim{\lower.5ex\hbox{\ltsima}}
\def\gtsima{$\; \buildrel > \over \sim \;$}
\def\gsim{\lower.5ex\hbox{\gtsima}}
\def\be{\begin{equation}}
\def\ee{\end{equation}}
\def\thv{\theta_{\rm{o}}}
\def\th0{\theta_{\rm{j}}}

\begin{document}

\title[Afterglow polarisation] 
{The polarisation of afterglow emission reveals GRBs jet structure}

\author[Rossi et al.]
{Elena M. Rossi$^1$, Davide Lazzati$^1$, Jay D. Salmonson $^2$ 
\& Gabriele Ghisellini$^3$ \\
$^1$Institute of Astronomy, University of Cambridge, Madingley Road,
Cambridge CB3 0HA, England \\ 
$^2$Lawerence Livermore National laboratory, L-038, P.O. Box 808, 
Livermore, CA, 94551 \\ 
$^3$Osservatorio Astronomico di Brera, Via E. Bianchi 46, I-23807 Merate, 
Italia \\
{\tt e-mail: emr,lazzati@ast.cam.ac.uk, salmonson1@llnl.gov, 
gabriele@merate.mi.astro.it}}

\maketitle

\begin{abstract}
 We numerically compute light and polarisation curves of
$\gamma$-ray burst afterglows for various configurations of the jet
luminosity structure and for different dynamical evolutions. We
especially consider the standard homogeneous ``top hat'' jet and the
``universal structured jet'' with power-law wings.  We also
investigate a possible more physical variation of the ``top hat''
model: the ``Gaussian jet''. The polarisation curves for the last two
jet types are shown here for the first time together with the
computation of $X$-ray and radio polarised fluxes.  We show that the
lightcurves of the total flux from these configurations are very
similar to each other, and therefore only very high quality data could
allow us to pin down the underlying jet structure. We demonstrate
instead that polarisation curves are a powerful means to solve the jet
structure, since the predicted behaviour of polarisation and its
position angle at times around the jet break are very different if not
opposite.  We conclude that the afterglow polarisation measurements
provide clear footprints of any outflow energy distribution (unlike
the lightcurves of the total flux) and the joint analysis of the total
and polarised flux should reveal GRBs jet structure.

\end{abstract}


\section{Introduction}

\begin{figure}
\psfig{file=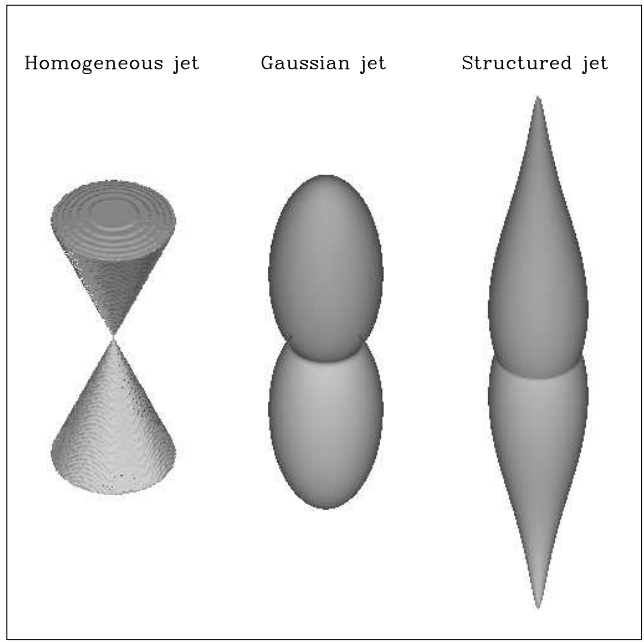,width=\columnwidth}
\caption{ Cartoon of the three jet configurations discussed in this 
paper. The figure shows the energy per unit solid angle of the jets
logaritmically scaled. The scale is different for each jet: 
it has been chosen in order to visually emphasize the characteristics
of each configuration.
\label{fig:cartoon}}
\end{figure}

After the discovery of a significant, albeit small, degree of linear
polarisation of the afterglow optical flux of GRB 990510 (Covino et
al. 1999; Wijers et al. 1999), there have been several other
detections of linear polarisation in a number of afterglows (see
Covino et al. 2002 for review). Typically, the polarisation is
observed to be at the 1-3 percent level (but see Bersier et al. 2003
for GRB 020405), with constant or smoothly variable level and position
angle when associated with a relatively smooth lightcurve (e.g. GRB
020813, Gorosabel et al. 2004). When deviations from a smooth power-law
decay in the lightcurve are instead present, polarisation curves show
a certain degree of complexity (e.g. GRB 021004 Lazzati et al. 2003;
Rol et al. 2003; GRB 030329 Greiner et al. 2003).

In order to observe polarisation, some kind of asymmetry is needed:
this can be provided by patches of coherent magnetic field, as
suggested by Gruzinov \& Waxman (1999) and Gruzinov (1999).  In
addition, small regions in which the magnetic field has some degree of
order could be amplified by scintillation (Medvedev \& Loeb 1999), or
by gravitational mini-lensing (Loeb \& Perna 1998; Ioka \& Nakamura
2001).  In these cases the required degree of asymmetry is in the
structure of the magnetic field, and not in the overall geometry of
the fireball, which could even be spherically symmetric.

Recently Granot \& K\"onigl (2003) proposed that the required asymmetry
might be provided by an ordered magnetic field embedded in the
circum-burst material, possibly amplified when the shock propagates
in it, to reach values close to equipartition with the energy density
of the shocked material in the ``pulsar wind bubble" scenario 
(K\"onigl \& Granot 2002).

A different class of models postulates that the fireball is jetted
(i.e. the ejecta are collimated into a cone with half opening angle
$\theta_{jet}$).  In this case the observer likely sees the fireball
off-axis, since the probability to be exactly on-axis is vanishingly
small (Ghisellini \& Lazzati 1999; Sari 1999, hereafter GL99 and S99,
respectively).  When the fireball bulk Lorentz factor $\Gamma$ is
$\sim 1/ (\theta_{jet}-\theta_o)$ (where $\theta_o$ is the viewing
angle) the emitting surface starts to be asymmetrical with respect to
the line of sight. Moreover it is assumed that a magnetic
field is compressed in the plane normal to the motion, analogous to
what has been proposed for the jets of radio-loud active galactic
nuclei (Laing 1980).  Face-on observers would see a completely tangled
magnetic field, but edge-on observers (in the comoving frame) would
see an aligned field, and therefore would detect synchrotron polarised
radiation.  Since the fireball is moving with a high bulk Lorentz
factor, the edge-on comoving observer corresponds to an observer in
the lab located at an angle $\sim 1/\Gamma$.  According to this idea,
there is a tight link between the behaviour of the total and the
polarised flux as a function of time. The light curve of the polarised
flux has two maxima (corresponding to the presence of the two edges of
the jet) with a position angle switched by $90^\circ$.  The maxima
occur just before and after the achromatic ``jet'' break in the light
curve of the total flux.  A main assumption of this model is that, at
any given angle from the apex of the jet, the luminosity emitted per
unit solid angle along the jet axis and along the jet borders is the
same.  Let us call jets with this energy structure ``homogeneous" jets
(HJs) (see Fig.1).

It is possible, instead, that the radiated power (per unit solid
angle) along the jet axis is larger than what is emitted along the
``wings".  If the wing energy distribution is a power-law, we refer
to these configurations as ``structured" jets (SJs) (see Fig.1).  As Rossi,
Lazzati \& Rees (2002) (thereafter RLR02) and Zhang \& M\'esz\'aros (2002)
have demonstrated, if the luminosity per unit solid angle is
$L(\theta)\propto\theta^{-a}$ with $a$ close to 2, then observers
with a viewing angle $\theta_{\rm o}$ would see an achromatic jet
break when $\Gamma\sim 1/\theta_o$.  It is therefore possible that
all GRB jets are intrinsically alike, having the same total intrinsic
power and the same jet aperture angle: they appear different only
because they are viewed along different orientations.  If the jet were
uniform, instead, they should have a large variety of aperture angles,
to account for the different observed jet-break times (Frail et al. 2001).

As demonstrated analytically in RLR02 and more recently numerically by
Salmonson (2003) (thereafter S03), it is difficult, on the basis of
the observed light curve, to discriminate between homogeneous and
structured jets (see also Granot \& Kumar 2003).  However, as
described in this paper, the two models are markedly different in the
polarisation properties of the produced afterglow flux.  In both
models the polarisation is produced because different parts of the
emitting jet surfaces do not contribute equally to the observed flux.
In the homogeneous jet model this starts to occur when $1/\Gamma$
becomes of the order of $\theta_{\rm jet}-\thv$ (i.e. when the
emitting surface available to the observer ``touches" the near border
of the jet).  In the structured jet model, instead, the required
asymmetry is built-in in the assumption that the emission is a
function of $\theta$, so that the relevant emitting surface is never
completely symmetric for off-axis observers.

 Finally we consider a jet with a Gaussian luminosity distribution
(Zhang \& Meszaros 2002). This can be regarded as a more realistic
version of the sharp edged standard jet: the emission drops
exponentially outside the typical angular size ($\theta_c$), within
which it is roughly constant  (see Fig.1).  Let us call it the
``Gaussian jet'' (GJ).  It has been argued that this configuration can
accomodate a unified picture of GRBs and X-ray flashes.  The
underlying assumptions is the presence of an emission mechanism for
which the peak energy $E_p$ in the prompt emission is a decreasing
function of the angular distance from the jet axis. In this way X-ray
flashes would be the result of observing a GRBs jet at large
angles. According to Zhang et al. (2003) the GJ would reproduce the
observed correlation $E_p \propto E_{iso}^{1/2}$ 
(Amati et al., 2002), while under the same
assumption the universal SJ and the ``top hat'' jet would face severe
problems.  We notice here, however, that the above correlation is
still based on a very small database in the X-ray flashes regime
(only two X-ray flashes are included) and
should be confirmed by future observations.  As regards afterglow
properties, a GJ seen within the core produces lightcurves that are
similar (but with smoother breaks) to the HJ's ones (Granot \& Kumar
2003). The luminosity variation with angle gives, as in the case of a
SJ, a net polarisation without the need of edges and we show here that
its temporal behaviour is indeed different from both the SJ's and the
HJ's one.

The detailed analysis of the polarisation characteristics and their
connection with lightcurves in  the homogeneous jet, universal
structured jet and in the Gaussian jet models is the main goal of
this paper.  In Section 2 we described the numerical code we have
implemented to study these models.  Some analytical and
semi-analytical results have been derived by GL99 and S99, for a
non-spreading jet and for a sideway expanding jet respectively. The
simplified prescription assumed by S99 for the lateral expansion led
to prediction of a third peak in the light curve of the polarised flux for an
observer close to the border of the jet.  In addition, GL99 did not
consider, for simplicity, the effects of the different travel times of
photons produced in different regions of the fireball, while S99
considered this effect by representing the viewable region as a thin ring
centred around the line of sight: the ring has an angular size of
$~\Gamma^{-1}$ and a constant width with respect to the ring radius.
Our numerical approach allows us to include the effects of the different
photon travel time and to analyze and compare different
prescriptions for the side expansion of the fireball.

This paper is organised as follows. In \S 2 we present the model for
the jet and for the magnetic field. In \S 3 we show the results for a
homogeneous jet, in \S 4 those for a structured jet and  in \S 5
those for a Gaussian jet. The comparison and discussion can be found
in \S 6. Finally in \S 7 we derive and discuss our conclusions, adding
possible complications to the models.

\noindent
Throughout this paper the adopted cosmological parameters are $H_0=65$,
$\Omega_{\lambda}=0.7$ and $\Omega_{m}=0.3$.

\section{The Code}
\subsection{The Jet structure and dynamics}

In this paper we show results obtained with two different codes.  The
first one is fully described in S03 while the second one is discussed
in this section. The main difference between the two codes is in the
treatment of the sideway expansion and dynamics in the
non-relativistic phase.  In the following we only remind the reader of
the different assumptions adopted by S03 and refer the reader to the
paper for more details.

We assume that the energy released from the engine is  in
the form of two opposite jets. They are described by the following
distributions of initial Lorentz factor $\Gamma_0$ and energy per unit
solid angle $\epsilon$ with respect to the jet axis ($\theta=0$):
\be
\begin{array}{ll}
\epsilon(\theta)=\frac{\epsilon_{c}}{\left(1+ \left(\frac{\theta}{\theta_{c}}\right)^
{(\alpha_{\epsilon}\,\beta_{\epsilon})}\right)^{\frac{1}{\beta_{\epsilon}}}} 
& \;\;\;\theta \leq  \theta_{jet},
\end{array}
\label{eq:E}
\ee

\be
\begin{array}{ll}
\Gamma_0(\theta)=\frac{\Gamma_{c}}{\left(1+ \left(\frac{\theta}{\theta_{c}}\right)^
{(\alpha_{\Gamma}\,\beta_{\Gamma})}\right)^{\frac{1}{\beta_{\Gamma}}}} 
& \;\;\;\theta \leq  \theta_{jet},
\end{array}
\label{eq:G}
\ee

\noindent
where $\theta_{jet}$ is the jet opening angle, $\theta_{c}$ is the
core angular size, $\epsilon_{c}=\epsilon(0)$ and
$\Gamma_{c}=\Gamma(0)$. 
In the following, in order to make the
comparison with the homogeneous jet easier, we will use preferentially
the local isotropic equivalent energy, defined as
$E_{iso}(\theta)=4\,\pi\,\epsilon(\theta)$.  In Eqs.~\ref{eq:E}
and~\ref{eq:G} $\alpha_{\epsilon},\alpha_{\Gamma}$ controls the shape
of the energy and $\Gamma_0$ distributions in the wings, while
$\beta_{\epsilon},\beta_{\Gamma}$ controls the smoothness of the joint
between the jet core and its wings. 

If $\alpha_{\epsilon},\alpha_{\Gamma}=0$, equations~\ref{eq:E}
and~\ref{eq:G} describe the standard top hat model with sharp edges
(homogeneous jet).  If the observer line of sight is located within
the jet, the observer detects the GRB prompt phase and its GRB
afterglow (GA); if the viewing angle $\theta_o$ is larger then
$\theta_{jet}$, he observes what it is called an orphan afterglow,
an afterglow not preceded by the prompt $\gamma$-ray emission.

When $\alpha_{\epsilon}>0$, the code describes a structured jet; in
this case $\theta_{jet}$ is assumed to be always much larger then the
observer angle. In fact we consider here a boundless jet (the end of
the wings are so dim that are undetectable), in contrast to the sharp
edged homogeneous jet. If $\alpha_{\epsilon}=2$ and
$\beta_{\epsilon}\to\infty$, the structured jet is that described in
RLR02, while S03 adopts $\alpha_{\epsilon}=2$ and
$\beta_{\epsilon}=1$.  

 For the Gaussian jet we use instead 
\be
\epsilon(\theta)=\epsilon_0 \,\,e^{-\frac{\theta^{2}}{2\theta_c^{2}}}.
\label{eq:gaus}
\ee

For simplicity we assume axial symmetry.  Our initial Lorentz factor
distribution satisfies $1/\Gamma\le \theta$, therefore regions on the shock front with different $\Gamma_0$ and
energy are causally disconnected and they evolve independently until
$1/\Gamma=\theta$.  In the numerical simulations we assume
for all models a costant initial Lorentz factor across the jet, with
$\Gamma_0=10^{4}$. With this choice the lightcurves shown in this
paper (for observed times $\gsim 15$ min) are insensitive to the
initial $\Gamma$ distribution.  As a matter of fact, for any
$\Gamma_0(\theta)\gsim 50$, the fireball deceleration starts earlier 
($t_d \simeq 245 s
(E_{53}/n_0/\Gamma_{0,2}^{8})^{1/3}$)
than the smallest time of the figures  and afterwards the evolution
follows the BM self similar solution and consequently the shown 
afterglow properties are independent from the initial Lorentz factor.
 If
relativistic kinematic effects freeze out the lateral expansion or the
pressure gradient prevents mixing, the different parts of the flow are
virtually independent along their entire evolution. We allow each
point of spherical coordinates ($r,\theta,\phi$) to evolve
adiabatically and independently, as if it were part of a uniform jet with
$\epsilon=\epsilon(\theta)$, $\Gamma_0=\Gamma_0(\theta)$ and
semi-aperture angle $\theta$. Therefore, if the mixing of matter is
unimportant this treatment is correct at any time, otherwise it gives
an approximate solution for $1/\Gamma\gg\Delta \theta$.  Actually
numerical hydrodynamical simulations seem to suggest that
$\epsilon(\theta)$ does not vary appreciably with time until the
non-relativistic phase sets in (Kumar \& Granot 2003) thus supporting
our numerical approach.

The full set of equations that determine the
dynamics of each patch of the jet is:
\be
\Gamma=\frac {\sqrt{1+4\,\Gamma_0\,f+4\,f^{2}}-1}{2\,f},
\label{eq:gamma}
\ee

\noindent
(e.g. Panaitescu \& Kumar 2000, thereafter PK00) 
where the parameter $f$ (the ratio of the swept-up mass to the initial
fireball rest mass) is given by:
\be
f=\frac{1}{M_0(\theta)}\int_{0}^{r}{r^2\Omega(r)\,\rho(r)\,dr},
\label{eq:f1}
\ee
where $M_0$ is the rest mass of the two (symmetric) jets, 
$\Omega(r)=4\pi\,(1-\cos\theta(r))$ is their solid angle and $\rho(r)$
is the ambient medium matter density.
The evolution of the solid angle is described by:
\be
\frac {d\theta}{dr}=\cos^2{\theta}\,\,\frac{c_s(\theta,r)} {c \beta \Gamma r}.
\ee
For the comoving lateral velocity $c_s$ we tested three different
recipes.  First, we analyze a non-sideways expanding (NSE) jet with
\be
c_s=0.
\label{nse}
\ee
then a sideways expanding (SE) one, either with a constant comoving 
sound speed (Rhoads 1999)
\be
c_s\simeq c/\sqrt{3}
\label{se2}
\ee
or with a more accurate treatment, which takes into account the
behaviour of the sound speed as a function of the shock Lorentz factor:
\be
c_s=c\,\sqrt{\frac{\hat{\gamma}(\hat{\gamma}-1)(\Gamma-1)}{1+\hat{\gamma}\,(\Gamma-1)}}
\label{se1}
\ee
(Huang et al. 2000), where $\hat{\gamma}=\frac{4\,\Gamma+1}{3\Gamma}$.

In the code developed by S03 (see in particular S03 \S2), it is
assumed that both momentum and energy are conserved (Rhoads 1999):
\be
\Gamma=\frac{\Gamma_0+f} {\sqrt{1+2\,\Gamma_0\,f+f^2}}.
\label{eq:gammajay}
\ee
This affects mainly the temporal slope of the lightcurve in
the non-relativistic regime, since the jet does not follow the
Sedov-Taylor solution.  For the sideways expansion prescription S03
assumes that the lateral kinetic energy of the shock, in its radially
comoving frame, is a constant proportion of the radial kinetic energy:
\be 
R_k = (\gamma'_\perp - 1)/(\gamma - 1),
\label{eq:sejay}
\ee
where $R_k$ (using S03's notation) describes the shock efficiency to
convert radial kinetic energy $(\gamma - 1)$ in lateral kinetic energy
($\gamma'_\perp - 1$) and
\be
\Gamma = \gamma \gamma'_\perp,
\ee
The jet dynamics in the trans- and non-relativistic phase is uncertain.
If the two opposite jets do not merge ($\theta_{jet}$ is always
$<90\degr$) when the dynamics becomes non-relativistic, the radial
momentum should be conserved through all the evolution of the jet.
This possibility depends strongly on the initial opening angle and on
the assumed type of lateral expansion. For example, using the lateral
velocity assumed by S03, initially narrow jets will not merge, because
the sideways expansion, deriving its energy from the forward
expansion, is soon exhausted.  On the other hand, if, at late times,
lateral spreading causes the jets to homogenize and become effectively
spherically symmetric, then Eq.~\ref{eq:gamma} is correct.  For these
reasons and the presence in literature of both treatments we compare
and show in this paper results from both Eq.~\ref{eq:gamma} and
Eq.~\ref{eq:gammajay}.

\begin{figure}
\psfig{file=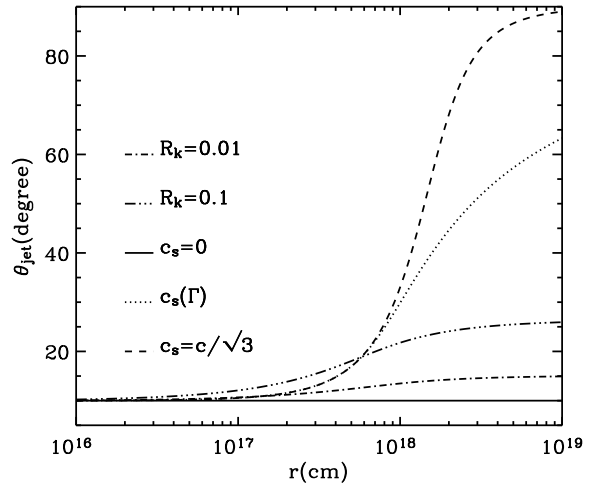,width=0.48\textwidth}
\caption{{The opening angle of an homogeneous jet as a function of 
the distance from the explosion site for four different lateral
velocities given by Eq.~\ref{nse}, Eq.~\ref{se2}, Eq.~\ref{se1} and
Eq.~\ref{eq:sejay}. The jet has an initial $\theta_{jet}$ of
$10^{\circ}$, $\Gamma_0=10^{4}$, $E_{iso}=10^{53}$ and the external
density is $n=10$ cm$^{-3}$.  See text for discussion.}
\label{fig:tetjr}}
\end{figure}

In Fig.~\ref{fig:tetjr} we show the opening angle of the jet vs. the
shock radius for all sideways velocity prescriptions.  The jet
spreads more efficiently when the simpler prescription
$c_s=c/\sqrt{3}$ is adopted while with Eq.~\ref{eq:sejay} the
expansion stalls out at fairly small angles ($\sim
\theta_{j}(0)+\arctan{\sqrt{R_k}}$), since the jet is radially
decelerating.  Eq.~\ref{eq:sejay} gives another interesting behaviour:
the jet begins to laterally expand earlier than $\theta_{jet}(r)\sim
1/\Gamma$; this is because $R_k$ = 0.01,0.1 correspond to very large
and supersonic initial lateral expansions.  A direct comparison
between the two sonic expansions (dotted and dashed lines) shows that
a variable sound velocity (Eq.~\ref{se1}) gives a final $\theta_{jet}$
smaller than Eq.~\ref{se2}.  Note however that this seems not to
affect appreciably the resulting lightcurve (Fig.~\ref{fig:omcomlc}).

\subsection{Spectrum and Luminosity}

In order to isolate the effects of the jet structure on the emission
and polarisation curves, we assume throughout this paper a constant
density environment $\rho(r)=m_p\,n$ ($m_p$ being the proton mass and
$n$ the number density). Polarisation curves derived here should be
therefore compared only to data from afterglow with smooth lightcurves
(Lazzati et al. 2003). Polarisation curves in a windy environment are
similar to ISM ones, but characterised by a slower evolution (Lazzati
et al. 2004).

The comoving intensity $I'$ is assumed to be given by the synchrotron
process only (Granot \& Sari 2002); we do not include inverse Compton
emission, because (for the adopted fiducial parameters) it can modify
the observed afterglow lightcurve and polarisation curves only in the
X-ray band after $\sim 10$ days (depending on the density, see
e.g. Sari \& Esin 2001).  Therefore our results are accurate up to the
X-ray band. We will comment on how the polarisation curves would be
modified, should IC emission be included (\S 3.3.4).

In the shocked matter the injected electron Lorentz factor is
\be
\gamma_m=\frac{m_p}{m_e}\,\epsilon_e \left(\Gamma-1\right),
\label{eq:gm}
\ee
where $m_e$ is the electron mass and $\epsilon_e$ is the fraction of
kinetic energy given to electrons; the Lorentz factor of the electrons
that cool radiatively on a timescale comparable to dynamical timescale
(i.e. time since the explosion) is
\be
\gamma_c=15\,\pi\,\frac{m_e\,c^{2}\,\sqrt{\Gamma^2-1}}{\sigma_{T}\,B^{2}\,r},
\label{eq:gc}
\ee
where $\sigma_{T}$ is the Thomson cross section and the comoving
magnetic field $B$ is given by:
\be
B=\sqrt{32\,\pi\,\epsilon_B\,m_p\,c^2\,n}\,\sqrt{\Gamma^2-1}.
\label{eq:B}
\ee

\noindent
where $\epsilon_B$ is the fraction of internal 
energy that goes to the magnetic field. 
Note that Eq.~\ref{eq:gc} and Eq.~\ref{eq:B} are different from those
given by PK00 that are not accurate in the trans
and sub-relativistic regimes. The corresponding synchrotron
frequencies are:
\be
\nu_{i}=0.25 \,\frac {e\,\gamma_{i}^2\,B}{m_e\,c},
\ee
where the constant (0.25) is calculated for an electron energy distribution 
index $p=2.5$ (PK00), $e$ is the electron charge,
 $i=m$ for the peak frequency and $i=c$ for the cooling
frequency.  The synchrotron self absorption frequency is given by:
\be
\nu_a=\nu_{i}\,\tau_{i}^{3/5},
\ee
where $i=m$ in the slow cooling regime and $i=c$ in the fast cooling
regime; $\tau_{i}=\frac{5\,e}{3}\,\frac{n\,r}{B\gamma_{i}^5}$ is the
optical thickness at $\nu_{i}$.  The comoving peak intensity
$I^\prime_p$ is
\be
I'_p=\frac {P'\,n\,r\,}{\nu_m},
\ee
where
$P'=\frac{4}{3}\,\sigma_{T}\,c\,\frac{B^{2}}{8\,\pi}\left(\gamma_m^{2}-1\right)$
is the total power emitted by relativistic electrons with Lorentz
factor $\gamma_m$ and isotropic distribution of pitch angles.  The
local observed luminosity is then computed through:
\be
dL(t,\theta,\phi)=I'\,\delta^{3}\, r^{2}\, \sin{\theta}\, d\theta\, d\phi
\label{eq:lum}
\ee
where $\delta=\frac{1}{\Gamma\,(1-\beta \cos{\tilde{\theta}})}$ is the
relativistic Doppler factor and $\tilde{\theta}$ is the angular
distance from the line of sight. The observed arrival time $t$ of
photons can be computed as follows:
\be
t=t_{lab}-\frac{r}{c \cos\tilde{\theta}},
\label{eq:sup}
\ee
where the time in the laboratory frame $t_{lab}$ at which the photons
were emitted is
\be
t_{lab}=\int \frac{dr}{\beta_{sh} c},   
\ee
where $\beta_{sh}\,c$ is the shock front speed.
The Lorentz factor $\Gamma_{sh}$ of this front is related to the
Lorentz factor $\Gamma$ of the shocked matter behind it
(Eq.~\ref{eq:gamma}) by
$\Gamma_{sh}=1+\sqrt{2}\,\left(\Gamma-1\right)$ (e.g. Sari 1997).

\subsection{The emitting volume}
\begin{figure}
\psfig{file=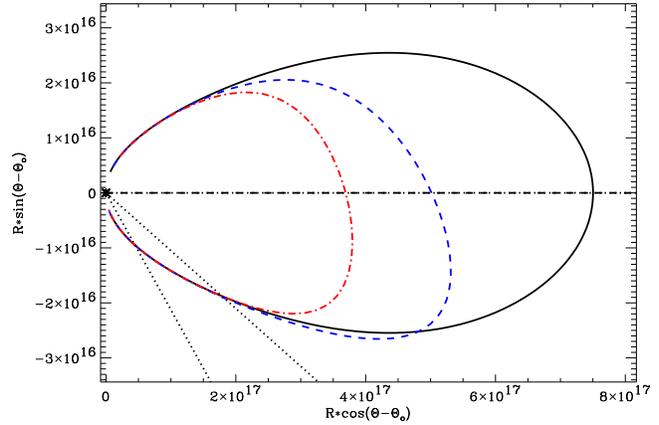,width=0.48\textwidth}
\caption {{The Equal Arrival Time Surfaces for a a SJ seen at
 $0^{\circ}$ (solid line), $6^{\circ}$ (dashed line) and 
$12^{\circ}$ (dot-dashed line) from the jet axis. 
 The straight dotted lines mark the position of the jet axis
and the horizontal dashed line is the line of sight.
For $\theta_o=0$ of course the dotted line overlaps the 
dashed line.
The jet parameters are: external density
$n=1$ cm$^{-3}$, core dimension $\theta_c/\theta_{jet}=0.1$,
$E_{c}=10^{54}$ erg, $\Gamma_c=10^4$, $\alpha_{\epsilon}=2$ and
$\alpha_{\Gamma}= 2$. EATSs clearly depend on the observer position:
each line of sight mimics an homogeneous jet with different
energy, Lorentz factor and $\theta_{jet}$. }
\label{fig:sur}}
\end{figure}

For a given time, Eq.~\ref{eq:sup} describes the locus of points from
which photons arrive simultaneously at the detector (equal arrival time
surfaces EATS).  The EATS for a  structured jet are
shown in Fig.~\ref{fig:sur},
where the observer is located to the right.  Unlike the homogeneous
case (e.g. Panaitescu \& M\'esz\'aros 1998; Granot, Piran \& Sari 1999), 
the EATS shape in the SJ depends on the viewing angle. This is
because, for relativistic effects, each line of sight mimics a
homogeneous jet with different parameters: $\theta_{jet}\simeq
2\,\theta_{o}$, $E_{iso}\simeq E_{iso}(\theta_o)$ and
$\Gamma_0\simeq\Gamma_0(\theta_{o})$ (RLR02).

The emission that the observer detects at time $t$ does not come only
from a thin layer near the EATS; in fact it comes from a sizable
fraction of the volume behind it whose width is $\sim
\frac{r}{2\,\Gamma_{sh}}$. Therefore, for each $t$, $\theta$ and
$\phi$, we integrate equation~\ref{eq:lum} also over this emitting
volume, using the Blandford \& Mckee (1976, hereafter BM) solution for
the Lorentz factor of the shocked gas behind the shock front (Granot,
Piran \& Sari 1999).  This calculation for the observed flux
is strictly valid for $\nu>\nu_a$, since we do not 
take into account self absorbing effects.

\subsection{Magnetic field configuration and linear polarisation}

The magnetic field configuration we adopt is obtained by
compressing, in one direction, a volume containing a random magnetic
field: it has some degree of alignment seen edge-on while it is still
completely tangled on small scales in the uncompressed plane.  This
could be the geometry of a magnetic field produced by the blastwave
that, sweeping up the external medium, could confine the field in the
sky plane. It may also be the natural configuration of shock generated
magnetic fields (Medvedev \& Loeb 1999).

Since the fireball is relativistic, the circle centered around the
line of sight, with angular aperture $\tilde{\theta}\simeq 1/\Gamma$,
is the region contributing the most to the observed emission and the
photons that reach the observer from the borders of that circle are
emitted at $\theta'=90^{\circ}$ in the comoving frame.  Therefore the
light coming from an angle $\tilde{\theta}\simeq1/\Gamma$ has the
maximum degree of polarisation $P_0$ while the light travelling along
the line of sight is unpolarised (see Fig.~1 in GL99). We assume
$P_0=60\%$. In the comoving frame this value decreases with the
angular distance ($\theta'$) to the line of sight as
\begin{equation}
P(\theta)=P_o \frac{\sin^2(\theta')}{1+\cos^2(\theta')},
\label{eq:pteta}
\end{equation}
(Laing 1980). The Lorentz transformations of angles give us the
relation between $\tilde{\theta}$ and $\theta'$:
\be
\cos{\theta'}=\frac{\cos{\tilde{\theta}}-\beta}{1-\beta\,\cos{\tilde{\theta}}}
\ee
\be
\sin{\theta'}=\Gamma \sin{\tilde{\theta}}\,\,\left(1+\beta\,\cos{\theta'}\right).
\ee

In terms of the Stokes parameters, $Q$, $U$, $V$, and local luminosity
$dL$ (equivalent to the Stokes parameter $I$) $P(\theta)$ can be
described as a vector with components:
\be
dQ(t,\theta,\phi)=P(\theta) dL(t,\theta,\phi) \cos(2\phi),
\label{eq:Q}
\ee
\be
dU(t,\theta,\phi)=P(\theta) dL(t,\theta,\phi) \sin(2\phi).
\label{eq:U}
\ee
and 
\be
V=0
\ee
Integrating Eq.~\ref{eq:U} and Eq.~\ref{eq:Q} over the EATS we obtain
the intensity of the observed total polarisation vector at a time $t$:
\begin{equation}
\mathcal{P}(t)=\frac{\sqrt{Q(t)^2+U(t)^2}}{L(t)},
\label{eq:P}
\end{equation}
which is actually the fraction of flux linearly polarised.  The
direction of the total polarisation vector is then
\begin{equation}
\phi(t)=0.5\,\arctan{\frac{U(t)}{Q(t)}}.
\end{equation}

\begin{figure}
\psfig{file=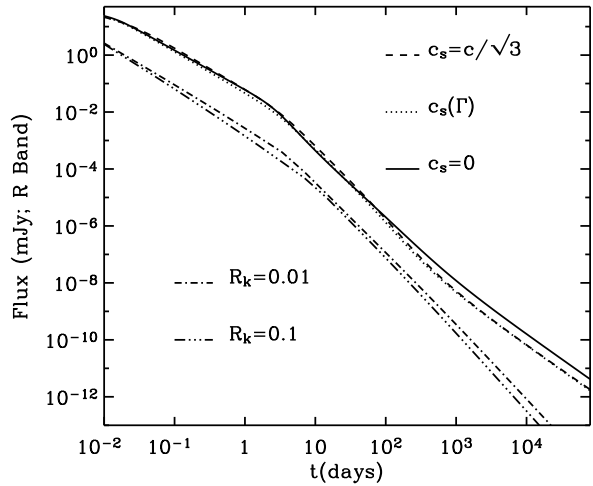,width=0.48\textwidth}
\caption{{Comparing the $R$ band ($7\times 10^{14} Hz$)
lightcurves resulting from different sideways 
expansions shown in Fig.~\ref{fig:tetjr}, where the jet parameters are
given. The shock parameters are: $p=2.5$, $\epsilon_e=0.01$ and $\epsilon_B=0.005$. An on-axis observer ($\theta_o=0$) has been assumed here.
 SE jets with constant (dashed line) and variable (dotted-line)
sonic lateral velocity give very similar lightcurves. SE jets with
supersonic lateral velocities (``dash-dot'' and ``dash-3 dots'' lines)
have been divided by a factor of $10$ for clarity. They
have more extreme spreading jet features (smoothness of the break,
late time break) and due to a different dynamics they do not follow
the Sedov-Taylor solution in the non-relativistic regime.}
\label{fig:omcomlc}}
\end{figure}

\begin{figure}
\psfig{file=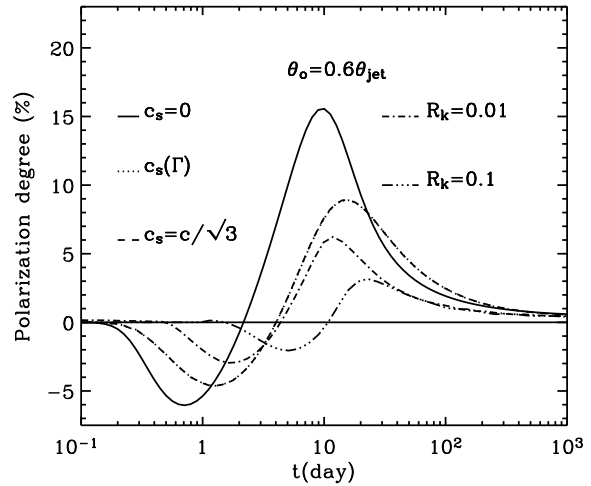,width=0.48\textwidth}
\caption{{ The same as Fig.~\ref{fig:omcomlc} for the polarisation curves.
SE jets with constant (dashed line) and variable (dotted-line) lateral
velocity give almost indistinguishable polarisation curves 
 (the second highest curve). 
Most of the time SE jets have a lower degree of polarisation than the NSE jet (solid line). 
Note that the larger  the lateral velocity, the lower  the polarisation.}
\label{fig:omcompo}}
\end{figure}

\begin{figure}
\psfig{file=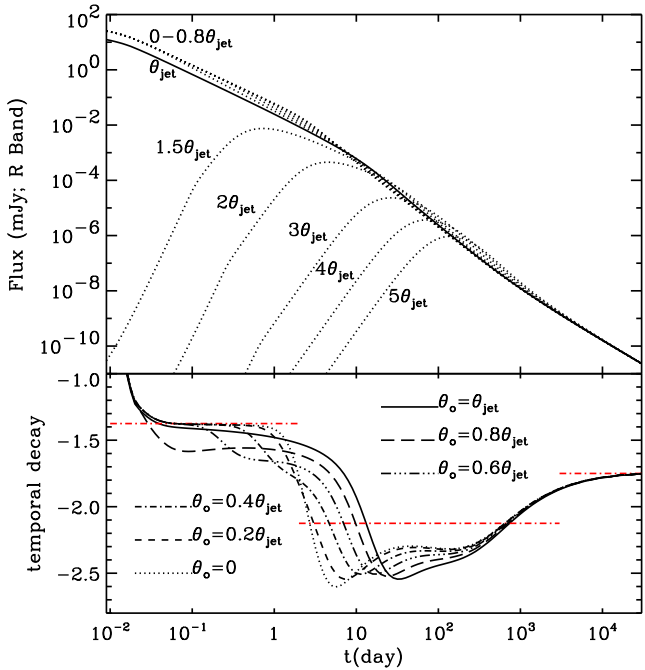,width=0.48\textwidth}
\caption{{Lightcurves (upper panel) and power-law 
indices as a function of time (lower panel) for a NSE homogeneous jet
($\alpha_{\epsilon}=\alpha_{\Gamma}$=0 and
$\beta_{\epsilon}=\beta_{\Gamma}=1$), seen from different viewing
angles.  The parameters are $\Gamma_0=10^{4}$, $E_{iso}=10^{53}$ erg,
$\theta_{jet}=10^{\circ}$, $n=10\,$cm$^{-3}$, $\epsilon_e=0.01$,
$\epsilon_B=0.005$, $p=2.5$, $\nu=7\times 10^{14}$ Hz and $z=1$.  For
these parameters the $R$ band is beyond the cooling frequency. The
curves corresponding to $\theta_{o}=0 \to \theta_{o}=0.8\theta_{jet}$
are shown but not labelled one by one for clarity. The black solid
line corresponds to $\theta_{o}=\theta_{jet}$, that shows a sort of
intermediate behavior between GRB afterglows
($\theta_{o}<\theta_{jet}$) and orphan afterglows
($\theta_{o}>\theta_{jet}$). The temporal decay for $GRB$ afterglows
is shown in the lower panel. The horizontal dash-dotted lines are the
standard analytically predicted slopes for the pre-break power-law
 $\alpha_1=-\frac{(3p-2)}{4}=-1.375$, the post-break power-law
$\alpha_2=\alpha_1-\frac{3}{4}=2.125$ and the non-relativistic phase
$\alpha_3=-\frac{(3p-4)}{2}=1.75$.  The
post break slope is steeper than predicted by a factor $\sim 1/4$.
Moreover as $\theta_{o}$ increases, the break becomes smoother (see
also Tab.~\ref{tab}), and the time of the turnover increases.  See
text for discussion.}
\label{fig:nse_lc}}
\end{figure}

\begin{figure}
\psfig{file=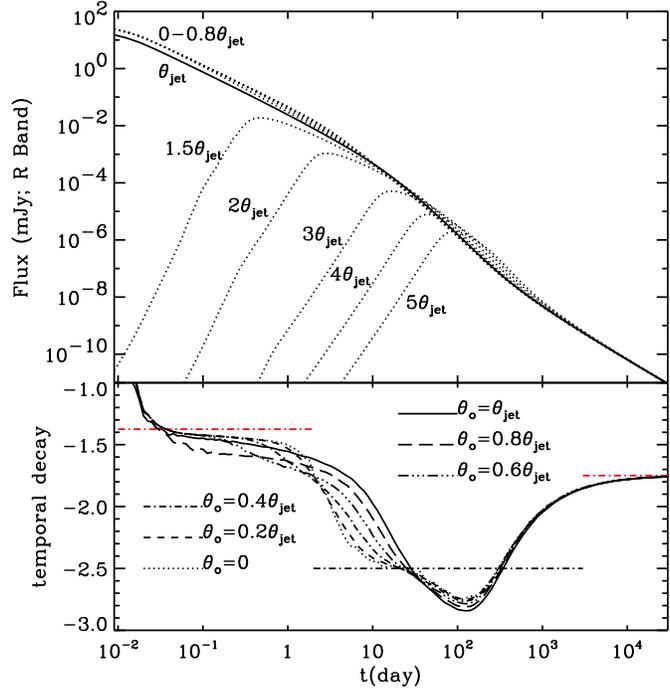,width=0.48\textwidth}
\caption{{Same as Fig.~\ref{fig:nse_lc}, but for a SE jet,  
with a comoving velocity given by Eq.~\ref{se1}.  The jet and shock
parameters are the same as Fig.~\ref{fig:nse_lc}.  Here the post break
slope given by the standard calculation is $\alpha_2=-p=2.5$, but again we
find $\alpha_2$ steeper by a factor of $\sim 1/4$, immediately after
the break. The lower panel shows that the time break increases with
$\theta_o$, as well as the smoothness of the break, though only mildly
(see also Tab.~\ref{tab}). See text for discussion.}
\label{fig:se_lc}}
\end{figure}

\section{Homogeneous jet: results}
 
In Fig.~\ref{fig:omcomlc} and Fig.~\ref{fig:omcompo} we compare the
light and polarisation curves resulting from different recipes for
the lateral expansion (\S 2.1).  The expansions given by Eq.~\ref{se2}
and Eq.~\ref{se1} produce very similar temporal behaviors; therefore
in the following we show results which are strictly valid for a
variable sound speed but the same discussion holds for
$c_s=c/\sqrt{3}$.  The lightcurves shown in Fig.~\ref{fig:omcomlc} are
all consistent with the same pre and post-break temporal slopes
for a  relativistic jet. However
the curves corresponding to a jet that obeys the conservation of
momentum (Eq.~\ref{eq:gammajay}, the dash-dotted and dash-3 dotted
lines) does not follow a Sedov-Taylor model ($\beta \propto r^{-3/5}$
and flux $F(t)\propto t^{-\frac{(3p-4)}{2}}$, for $\nu>\nu_c$) 
in the non-relativistic regime
but $\beta  \propto r^{-3}$ and  $F(t)\propto t^{-43/16}$.  Then in this
latter case the flux falls off more rapidly while in the former case
the lightcurve tends to flatten. We anticipate 
 that in both cases the
non-relativistic slopes do not depend on the viewing angle but only on
the spectrum, contrary to the temporal slopes in the relativistic
regime (see also Fig.~\ref{fig:nse_lc},~\ref{fig:se_lc}).

Fig.~\ref{fig:omcomlc} and Fig.~\ref{fig:omcompo} also show that a SE
jet has
\begin{itemize}
\item a smoother break in the lightcurve,
\item  a later time break,
\item  smaller polarisation peaks,
\end{itemize}
compared to a NSE jet and these characteristics are even more evident
for jets with a supersonic sideways expansion.

All these features (among others) are discussed in more details in the
following, comparing a NSE jet with a jet undergoing a sideways
expansion given by Eq.~\ref{se1}.  For a jet evolution following
Eq.~\ref{eq:sejay} we refer the reader to S03, where a more complete
discussion is given.

\subsection{Lightcurve: the shape of the break}

The R band lightcurve for a NSE and for a SE homogeneous jet are shown
in Fig.~\ref{fig:nse_lc} and Fig.~\ref{fig:se_lc} respectively.  They
show interesting features, some of which have never been discussed
before.  In the lower panels the temporal index $\alpha$ (defined as
$F(t)\propto t^{\alpha}$) is plotted versus time. We call
$\alpha_1$,$\alpha_2$ and $\alpha_3$ the pre-break, the post-break and
the non-relativistic slope respectively.  The horizontal dot-dashed
lines show the expected slopes from on-axis standard calculations.

\subsubsection{$\alpha_2$}
The ``standard'' post-break slopes (central dot-dashed lines in
Fig.~\ref{fig:nse_lc} and Fig.~\ref{fig:se_lc}, lower panels) are
calculated considering the loss of the emitting area from a spherical
blast wave $(1-\cos\Gamma^{-1})$ to a conical one
$(1-\cos\theta_{jet})$. For the SE jet we also consider the more rapid
deceleration due to the increase of the shock front surface (Rhoads
1999; Sari, Piran \& Halpern 1999). In both cases the surface
brightness is supposed to be the same across the surface of the jet at
all times. For $\nu>\nu_c$ the
expected breaks are $\Delta \alpha_{nse}=\frac{3}{4}$ for a NSE
jet and $\Delta\alpha_{se}=\frac{(p+2)}{4}=1.125$ for a SE one.

In fact we find that the flux after the break falls off more rapidly
than expected, and this effect is more evident as the electron energy
distribution $N(\gamma)\propto\gamma^{-p}$ becomes steeper. This can
be understood by taking into account the effect of EATS.  When
$r/\Gamma<r\,\theta_{jet}$ the visible area can be
schematized as a bright ring with radius $r/\Gamma$ and width $\Delta$
($20 \%-40\%$ of the whole area) that surrounds a dimmer uniformly
emitting surface (e.g. Granot et al. 1999).  When
$r/\Gamma>r\,\theta_{jet}+\Delta$ only the dimmer emitting surface
 is visible. As a consequence after the break the deficit in
the observed flux is bigger and $\alpha_2$ is steeper than considering,
at any time, a uniform emitting surface. The effect is more pronounced
for higher values of $p$ since the surface brightness contrast between
the centre and the ring increases. 
When $r/\Gamma \gg r\,\theta_{jet}+\Delta$ the lightcurve 
decay index should tend to the standard slope, because the jet surface emits
almost homogeneously; in fact only in the case of a very narrow jet
($\theta_{jet}<1^{\circ}$) this asymptotic slope is reached: wider jets
lightcurves break later and they enter the trans-relativistic phase
soon after the break, tending towards $\alpha_3$.

\subsubsection{$\alpha_1$}
Similarly the pre-break slope presents deviations from the standard
calculations for a spherical symmetrical blastwave ($\alpha_1=\frac
{-(3p-2)}{4}$: Fig.~\ref{fig:nse_lc} and Fig.~\ref{fig:se_lc}, lower
panels, horizontal line on the left). If the jet is very narrow (less
than few degrees) the lightcurve will not exhibit (depending on
$\Gamma_0$) this first branch. For wider jets (as in our examples) the
value $\frac {-(3p-2)}{4}$ is strictly followed only in the case of a
NSE jet for observers around the line of sight.
\subsubsection{$\alpha_3$}
On the other hand the non-relativistic regime offers a lightcurve
branch slope independent of the spreading and the viewing
angle. This holds for any frequency band. However, uncertainties in
the dynamics during the trans-relativistic phase (see \S 2.1) hamper
the possibility to derive the electron distribution
 power-law index even at late stages.

\subsection{ Lightcurve: the time of the break}
The lower panels of Fig.~\ref{fig:nse_lc} and Fig.~\ref{fig:se_lc}
show, in addition, how the jet-break and the temporal behaviour around
it change with the off-axis angle.  As the viewing angle
increases the transition between $\alpha_1$ and $\alpha_2$ is {\it
smoother} (GL99) and the break time (when $\alpha\simeq
\frac{(\alpha_1+\alpha_2)}{2}$) is {\it retarded}.

To obtain more quantitative results we fit the lightcurves $F(t)$ with
a smoothly joined broken power law (SBP),
\be
F(t)=\frac{2\,F_b}{\left[\left(\frac{t}{t_b}\right)^{\alpha_1\,s}+
\left(\frac{t}{t_b}\right)^{\alpha_2\,s}\right]^{1/s}},
\label{eq:fit}
\ee
(Beuermann et al. 1999), where $s$ is the smoothness parameters and
$F_b$ the normalisation ($F_b=F(t_b)$ for $s=1$).  The smoothness
parameters is a measure of the shape of the lightcurve around the
break: the lower its value the smoother the two asymptotic slopes are
joined together over the break.  The fit is performed over the
range\footnote{The break time $t_b$ is found iteratively by adjusting
the fitting interval to the one specified and re-performing the fit
until convergence.} $0.01\,t_b\le{}t\le100\,t_b$ and assigning an
uncertainty of $10\%$ to each point.

The results are summarised in Tab.~\ref{tab} where $t_b$ and $s$ are
given as a function of $\theta_o/\theta_{jet}$.  For  NSE jet lightcurves,
the fit gives break times that range from $1.86$ days for a on-axis
observer to $5.7$ days for $\theta_o/\theta_{jet}=0.8$, while for a SE jet
$t_b=3.57$ days for $\theta_o=0$ and $t_b=7.44$ days for
$\theta_o/\theta_{jet}=0.8$. However changing the
time interval of the fitting can result in a variation of the
estimated $t_b$ of a factor of 2, while the positive correlation between
$t_b$ and $\theta_o$ generally holds. Tab.~\ref{tab} shows also that
$s$ decreases for larger $\theta_{o}$ and the effect is greater when
lateral expansion is not dynamically important (see also Fig. 4 in
GL99). The bottom line of this discussion is that the break time is an
ill-defined quantity, since the function generally used for data fits is
only a rough approximation to the real shape of the afterglow
lightcurve. Systematic uncertainties on the measure should be
considered, especially when the break time is used to infer the
opening angle of the jet to derive the beam-corrected energy output of
GRBs (e.g. Berger et al. 2003).

\begin{figure}
\psfig{file=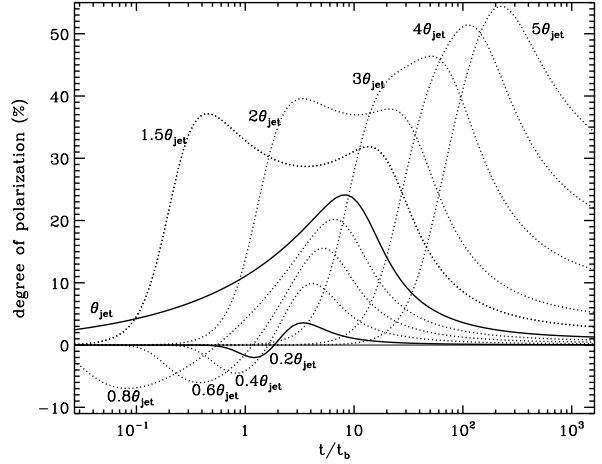,width=0.48\textwidth}
\caption{{Polarisation curves for a homogeneous NSE jet, corresponding
to the lightcurves shown in Fig.~\ref{fig:nse_lc}. The $x$ axis is the
time since the trigger divided by the on-axis lightcurve time break
$t_{b}=1.86$ days.  These curves may have to be rescaled by a factor
$<1$, since $P_0=60\%$ is taken arbitrary.  For $P<0$ the vector is in
the plane containing the line of sight and the jet axis, while for
$P>0$ is rotated of $90^{\circ}$. The black solid lines correspond to
$\theta_{o}=\theta_{jet}$ and $\theta_{o}=0.2\theta_{jet}$. GRB
afterglows always show two peaks, with the second one higher than the
first, and with polarisation angles rotated by $90^{\circ}$. Orphan afterglows' polarisation
curves have two peaks, with the same angle, for
$\theta_{o}<3\theta_{jet}$ that eventually merge into a single maximum
for larger viewing angles; the polarisation at the peak grows with
$\theta_{o}$. The peak polarisation for an orphan afterglow can
therefore be a factor $\sim 2.5$ larger than what it is expected at
$\theta_{o}=\theta_{jet}$.}
\label{fig:nse_po}}
\end{figure}

\begin{figure}
\psfig{file=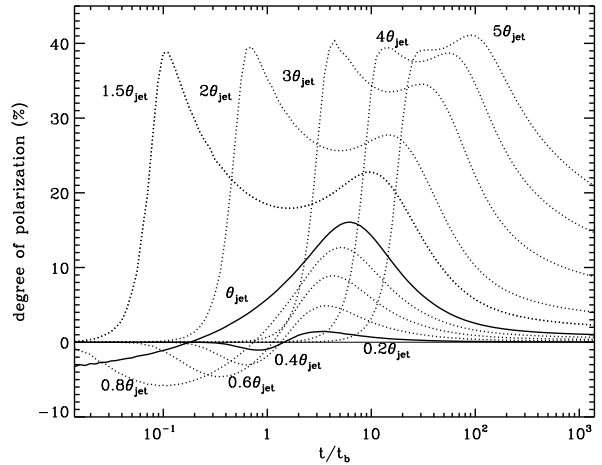,width=0.48\textwidth}
\caption{{Same as Fig.~\ref{fig:nse_po} but for a  homogeneous SE jet. 
The corresponding lightcurves are shown in Fig.~\ref{fig:se_lc}. The
$x$ axis is the time since the trigger divided by the on-axis
lightcurve time break $t_{b}=3.57$ days. All the main features are the
same as for the $NSE$ jet but the expected polarisation values are lower
and the two maxima in the orphan afterglows curve merge for
$\theta\simeq 6\theta_{jet}$, where the maximum value of polarisation
is reached. Then the polarisation at the peak decreases.
The polarisation for an orphan afterglow can therefore be a factor
$\sim 2.6$ larger than what it is expected at $\theta_{o}=\theta_{jet}$.}
\label{fig:se_po}}
\end{figure}

\subsection{Polarisation curves}
The polarisation curves we obtain are given in Fig.~\ref{fig:nse_po}
and Fig.~\ref{fig:se_po}.
\subsubsection{GRB afterglows}
Homogeneous GRB afterglow polarisation curves
($\theta_o/\theta_{jet}\le 1$) always show two peaks, with the second
higher then the first one, with pitch angles rotated by
$90^{\circ}$. For $\theta_o/\theta_{jet} \gsim 0.6$ this change in
the direction of the polarisation vector ($P=0$) happens before the
break-time measured by an on-axis observer, while for smaller angles
it happens slightly after.
\subsubsection{Orphan  afterglows}
The polarisation curve of orphan afterglows ($\theta_o/\theta_{jet}\ge 1$)
has two peaks, with the same position angle, that eventually merge in
a single maximum.  For a NSE jet the polarisation at the peak grows
with $\theta_{o}$, tending towards $P_{0}$ (Fig.~\ref{fig:nse_po}),
while with sideways expansion the peak value reaches a maximum around
$\theta_{o}\sim 7
\theta_{jet}$ and then it slowly decreases; in both cases the
polarisation peak for an orphan afterglow can be a factor $\sim
2.5-2.6$ larger than what it is expected at $\theta_{o}=\theta_{jet}$. 
\subsubsection{Comparison with previous results}
The polarisation curves for a NSE jet with $\theta_o<\theta_{jet}$ have
been previously published by GL99, which did not consider EATS.  Their
curves have our very same temporal behaviour, but their polarisation
peaks are lower than ours by a factor from 2 to 4, depending on the
viewing angle. This is the result of adding EATS to the computation.
In this case the received intensity, at any given time, peaks at an
angular distance $1/\Gamma$ from the line of sight, just where the
linear polarisation is maximised (see Eq.~\ref{eq:pteta},
Eq.~\ref{eq:U} and Eq.~\ref{eq:Q}).  As a result, the total expected
polarisation is higher than for a homogeneously emitting surface.
S99 and Granot \& K\"onigl (2003) have instead explored the
polarisation for a spreading jet. The first author uses a simplified
model in which the opening angle does not change until
$\frac{1}{\Gamma}<\theta_{jet}$ and then $\theta_{jet}$ increases as
$1/\Gamma$; as a consequence he expects a third polarisation peak to
appear at later times for large off-axis angles, while for small
viewing angles only one peak should be visible.  With our complete
calculation of the evolution of the opening angle of the jet we do not
obtain either of these effects (see Fig.~\ref{fig:se_po}). The reason
is that the visible area ($1/\Gamma$) crosses both the nearest and the
farthest edge of the jet for any off-axis angle and then $1/\Gamma$
remains always greater than $\theta_{jet}$. Therefore for a HJ, afterglow
polarisation curves \emph{ always have only two peaks with orthogonal
polarisation angles}, for all the sideways expansion models considered in
this paper.  Granot et al. (2002) have extended the computation also
to orphan afterglows.  They obtain
\vspace{-0.25cm}
\begin{itemize}
\item[i)] higher value of polarisation for $GRB$ afterglows compared to ours
with the first peak always greater then the second one (unlike what our
curves show); 
\item[ii)] for $\theta_{o}/\theta_{jet}=0.25$ only one peak is visible;
\item[iii)] an orphan afterglow peak can have a polarisation degree 
which is larger than what observed
by an on-axis observer, but only by a factor 2.
\end{itemize}
These effects are due to an error in their program, pointed out
recently by the authors themselves (see Granot \& K\"onigl 2003). The
results of their corrected code are in general agreement with ours
(Granot private communication).
\vspace{-0.25cm}
\begin{table}
\centerline{\begin{tabular}{|c|cc|cc|}
\hline
& \multicolumn{2}{c}{NSE}& \multicolumn{2}{c}{ SE} \\
\hline
$\theta_o/\theta_{jet}$ & $\frac{t_b}{t_b(0)} $ & $s$ &  $\frac{t_b}{t_b(0)} $ & $s$  \\ \hline
0 & 1 & 8.36 & 1 & 1.07 \\
0.2 & 1.01 & 4.34 & 1.06 & 0.96 \\
0.4 & 1.12 & 1.79 & 1.25 & 0.76 \\
0.6 & 1.67 & 0.86 & 1.73 & 0.55 \\
0.8 & 3.07 & 0.60 & 2.09 & 0.55 \\
\hline
\end{tabular}}
\caption{ This table summarises some results from 
the modelling with a SBP (Eq.~\ref{eq:fit}) of the lightcurves shown in
Fig.~\ref{fig:nse_lc} ($NSE$ jet) and Fig.~\ref{fig:se_lc} ($SE$ jet).
The break time $t_b$ is given as a multiple of the $t_b$ fitted by an
on-axis observer [$t_b(0)$]. Note that the time at which the curve
changes slope is postponed as the angular distance from the jet axis
increases.  The smoothness parameter $s$ of the break in a
lightcurve is a measure of the break shape: the higher its value,
the sharper is the transition between the two asymptotic slopes. The
table shows that $s$ decreases with the off-axis angle. This
effect is more evident for a $NSE$ jet.}
\label{tab}
\end{table}

\begin{figure}
\psfig{file=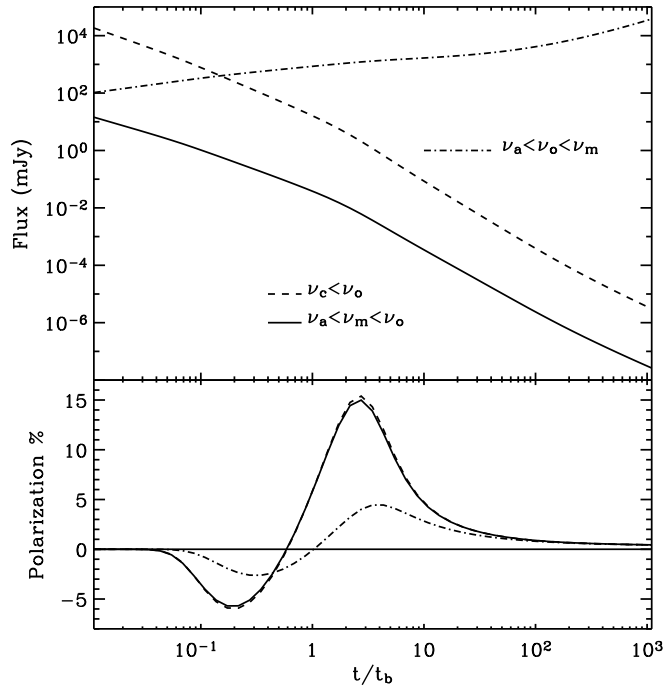,width=0.48\textwidth}
\caption{{ Lightcurves (upper panel) and polarisation curves (lower panel) 
for a NSE jet seen at $\theta_{o}=0.6\theta_{jet}$ in three different
spectral ranges: $F(\nu)\propto\nu^{1/3}$ (dot-dashed line, usually in
the radio) $F(\nu)\propto\nu^{-(p-1)/2}$ (usually in the optical,
solid line) and $F(\nu)\propto\nu^{-p/2}$ (usually in the X-rays,
dashed line).The jet and shock parameters are the same as
Fig.~\ref{fig:nse_lc}. $t_{b}=3.66$ days.  The X-ray lightcurves has
been rescaled by a factor $10^3$ for clarity.}
\label{fig:xr}}
\end{figure}
\subsubsection{Multiwaveband polarisation}
Polarisation due to synchrotron emission is in principle present at
any wavelength. In Fig.~\ref{fig:xr} we show the light and
polarisation curves for the same GRB afterglow observed at a frequency
$\nu_o$ in three different spectral branches: $\nu_a<\nu_o<\nu_m$,
$\nu_m<\nu_o<\nu_c$ and $\nu_c<\nu_o$, for an observer at
$\frac{\theta_o}{\theta_{jet}}=0.6$. In the following we refer to them as
the ``radio'' branch, the ``optical'' branch and the $X$-ray branch
respectively, since each waveband usually stays on that particular
branch for most of the afterglow evolution (depending on fireball and
shock parameters).

Fig.~\ref{fig:xr} shows that while the polarisation curves in the $R$
and X-ray bands are very similar, the radio curve has a significantly
lower degree of polarisation.  Disticnt from the other lightcurves,
the radio flux increases with time before the break ($F_{\nu}\propto
t^{1/2}$). It should fall asymptotically as $F_{\nu}\propto t^{-1/3}$
afterwards but the trans-relativistic phase sets in and eventually the
flux rises again ($F_{\nu}\propto t^{8/5}$) following the
non-relativistic temporal slope (Frail, Waxman \& Kulkarni 2000). 

We note that generally a spectral transition occurs in the radio band
(8.5 GHz) at late times, when the peak frequency becomes smaller than
the observed radio band ($\nu_a<\nu_m<\nu_o$).  This causes a third
peak to appear in the radio polarisation curve, since in the radio
curve joins the optical one; on the other hand the radio lightcurve
undertakes a spectral break
 and eventually follows the optical lightcurve temporal decay
($F_{\nu}\propto t^{-3(p-1)/4}$ before
 the jet break and $F_{\nu}\propto t^{-3 p/4}$ after).
In our calculations for the radio polarisation we do not take into
account the effects of Faraday rotation: as long as $\Gamma>2$ this
effect should be negligible, since the electrons are all highly
relativistic, with $\gamma_m>100$.  In the trans-relativistic phase
its effect can be sizable and it could even wash out the
intrinsic small degree of radio polarisation.

The $X$-ray polarisation curve can be affected by the contribution of
the inverse Compton flux; in particular after $10-15$ days it may
depolarise the synchrotron flux, depending on the external density.
However faint fluxes are usually observed at this epoch, too faint for
detection by any X-ray polarimeter conceivable in the near future. In
fact, detection of polarisation in the $X$-ray band will probably be
possible only within few hours from the trigger, when the synchrotron
flux positively dominates the inverse Compton emission.

\subsection{Summary}

In this section we have described separately lightcurves and
polarisation curves for homogeneous jets.  
The variation of viewing angle however affects
both quantities.  As the off-axis angle increases the break time increases
and since the minimum in the polarisation curves occurs around
$t_{b}(\theta_o)$, it is delayed as well.  Moreover the degree of
polarisation increases at the peaks while the break shape becomes
smoother.  These joint characteristics should, in principle, be
observable and be helpful for testing the model.

\section{Structured jet: results}
As a general result, our more sophisticated simulations confirm all the
features we described in RLR02 for the lightcurve of a SJ: it is very
similar to the lightcurve of an HJ seen on-axis with same energy per
solid angle and $\theta_{jet}=2\theta_o$.  On the other hand, we show that
the polarisation curves for a SJ present key-features that allow us to
spot the underlying jet structure.

\subsection{Lightcurves}

\begin{figure}
\psfig{file=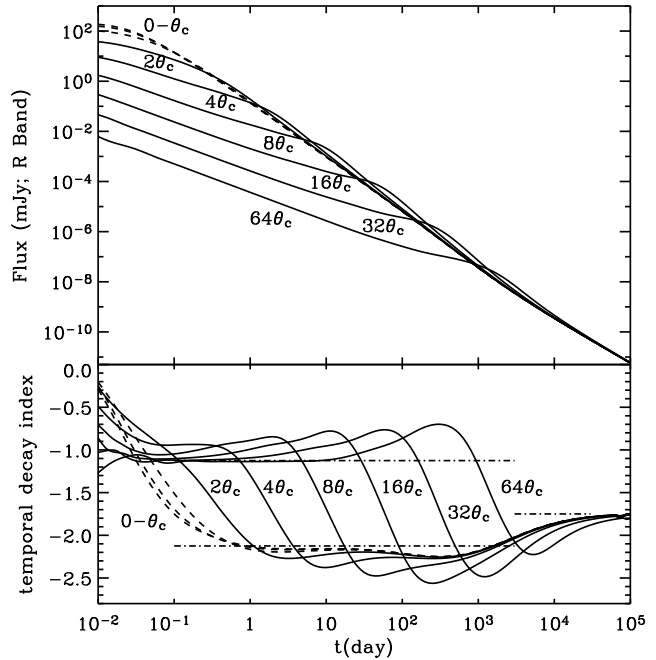,width=0.48\textwidth}
\caption{{ Lightcurves in the $R$ band and temporal index for a structured 
NSE jet with parameters: $E_{c}=2 \times 10^{54}$ erg,
$\alpha_{\epsilon}=2$, $\beta_{\epsilon}=1$, $\beta_{\Gamma}=0$,
$\Gamma_{c}=2\times10^{4}$, $\theta_c=1^{\circ}$,
$\theta_{jet}=90^{\circ}$, $\epsilon_e=0.01$, $\epsilon_B=0.005$ and
$n=1$ cm$^{-3}$. The viewing angle $\theta_o$ is indicated in the
figure. Before the break the $R$ band is below $\nu_c$, after the break
it is above. Thus the asymptotic slopes are:
$\alpha_1=-\frac{3}{4}(p-1)=-1.125$, $\alpha_2=-(3p+1)/4=-2.125$ and
$\alpha_3=-(3p-4)/2=-1.75$. }
\label{fig:nseinlc}}
\end{figure}

\begin{figure}
\psfig{file=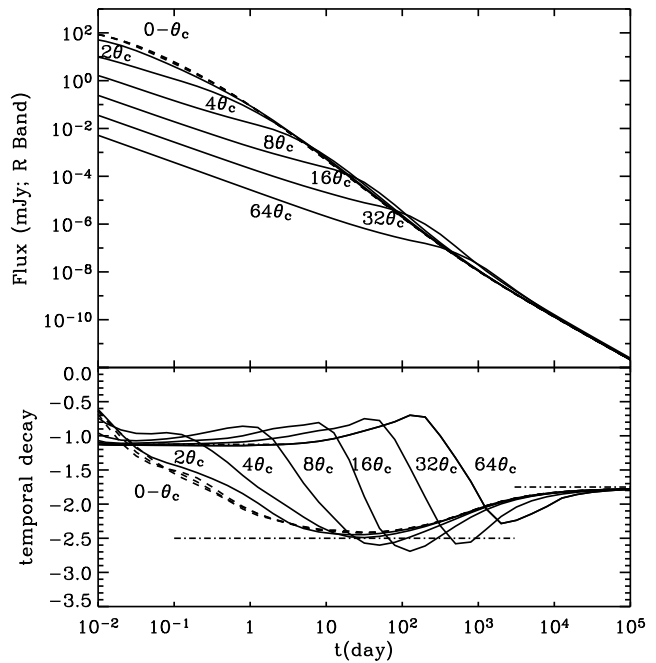,width=0.48\textwidth}
\caption{{The same as Fig.~\ref{fig:nseinlc} but for a structured SE jet
with a comoving sideways velocity given by Eq.~\ref{se1}. All the
other parameters are the same as Fig.~\ref{fig:nseinlc}. 
Here $\alpha_2=-p=-2.5$.}
\label{fig:seinlc}}
\end{figure}

The temporal index $\alpha$ of a SJ lightcurve is shown in the
lower panels of Fig.~\ref{fig:nseinlc} and Fig.~\ref{fig:seinlc} for 
$\alpha_{\epsilon}$=2, $\beta_{\epsilon}$=1 and  $\beta_{\Gamma}$=0.
Similar to the HJ, a SJ's lightcurve for $\theta_o>$ few degrees can be
approximated as a broken power-law; moreover the same asymptotic slopes
($\alpha_1$, $\alpha_2$, $\alpha_3$) for an {\it homogeneous} jet (see
\S 3.1) describe as well the temporal behaviour of a SJ's lightcurve.

It can be noticed that for viewing angles within the core
($\theta_o\le \theta_c$) the observed lightcurves are similar to those
expected from very narrow HJs: the break happens so soon that the
first power-law branch is missing while after the break there is
enough time to reach $\alpha_2$ and follow that slope until the
non-relativistic transition sets in. For larger viewing angles the
situation is quite the reverse: as the $\theta_o$ increases the
lightcurves have time to follow more strictly the standard value for
$\alpha_1$ but there is less and less time to reach $\alpha_2$ before
entering the sub-relativistic phase. This is the same behaviour
observed in HJs lightcurve as the jet opening angle increases.
Another point of similarity with HJ's lightcurves is that (as
discussed in \S 3.1.1 for HJs) the flux falls off more rapidly after
the break than expected by standard calculations: this effects
increases with the off-axis angle. Distinct from the HJ, the SJ
lightcurves also present a flattening before the break that increases
with the off-axis viewing angle. The deviations from the standard
slopes before and after the break are of the same order.

We can conclude that HJ's and SJ's lightcurves can be described only
roughly by broken power-laws and certainly only for a specific range
of viewing angles.

\subsubsection{The shape of the break}
To quantify the sharpness of the break we again fit the simulated
lightcurves with Eq.~\ref{eq:fit} and the results are summarised in
Table ~\ref{tab2}.  The smoothness parameter $s$ increases and then
saturates around $\theta_o/\theta_c=8$, where the fit yields an
extremely sharp break (the best fit is actually with a simple broken
power-law rather than a SBP). A $NSE$ jet has sharper jet breaks
compared to a $SE$ jet, but the relation between $s$ and the off-axis
angle is similar.  This relation is actually different from that
discussed for the HJ (\S 3.2), where the break becomes smoother and
smoother as $\theta_o$ increases.

\subsubsection{The time of the break}
The similarity between HJ and SJ lightcurves becomes more quantitative
if we measure the break-time as a function of the viewing angle.  The
break time $t_b$ increases with the viewing angle and the general
trend can be fitted by a power law: $$t_b\propto (\theta_o/\theta_c)^2 $$
 (see Tab.2).  We underline that
$t_b\propto \theta_o^2$ is predicted by the structured model with
$\alpha_{\epsilon}$=2 and by the HJ model with constant total energy
(once $\theta_o$ is replaced  by $\theta_{jet}$).

One difference between the HJ and the SJ is that, for the same set of
parameters, the SJ break time equals that of the HJ if $\theta_{jet} \sim
2\,\theta_o$ (RLR02) (see also Fig.~\ref{fig:inomcom}).  This factor
comes from the two different origins of the break in the lightcurve:
the HJ simply breaks when the edge of the jet comes into view
($\Gamma\simeq \theta_{jet}^{-1}$), while the structured jet breaks
when $\Gamma_c \simeq \theta_o^{-1}$. The following simplified
calculation aims to show that a HJ (with the same parameters as a SJ
and $E_{iso}=E_{iso}(\theta_o)$) tends to have an earlier break time if
$\theta_{jet}=\theta_o$ and thus a larger opening angle (of the order
of $\sim 2 \theta_o$) is needed to have a reasonable agreement between
the two break times.  The break times ratio for a non spreading jet is
\begin{equation}
\frac{t_{jet,s}}{t_{jet,h}} = \biggl(\frac{\theta_o}{\theta_{jet}}\biggr)^{8/3}
\left(\frac{\theta_o}{\theta_c}\right)^{2/3},
\label{eq:trnse}
\end{equation}
\noindent
where $t_{jet,s}=\frac{R_{jet,c}}{c}\,(1-\beta\cos(\theta_o))$ is the
observed break time for SJ ($R_{jet,c}$ being the radius at which
$\Gamma_c=1/\theta_o$) and
$t_{jet,h}=\frac{R_{jet}}{2\,c}\theta_{jet}^{2}$ is the break time for
an HJ seen on-axis ($R_{jet}$ being the radius at which $\Gamma\sim
1/\theta_{jet}$).  In Eq.~\ref{eq:trnse} and below in
Eq.~\ref{eq:trse} we assume the same isotropic equivalent energy along
the line-of-sight and that $\Gamma_0$ does not depend on $\theta$; for
a spreading jet we get instead

\begin{equation}
\frac{t_{jet,s}}{t_{jet,h}}=\biggl(\frac{\theta_o}{\theta_{jet}}\biggr)^{8/3}
\left(1+\ln{\frac{\theta_o}{\theta_c}}\right),
\label{eq:trse}
\end{equation}

\noindent
where the logarithmic term takes into account the exponential
behaviour of $\Gamma_c$ for $\Gamma_c<\theta_c^{-1}$.  If we impose
then that the structured and homogeneous jets lightcurves break at the
same time Eq.~\ref{eq:trnse} gives 
$1.3\lsim \frac{\theta_{jet}}{\theta_{o}}\lsim 2.5$ and
Eq.~\ref{eq:trse} give
s $1.3 \lsim \frac{\theta_{jet}}{\theta_{o}}\lsim 1.8$
for $2 \lsim \frac{\theta_o}{\theta_{c}}\lsim 40$.

\subsection{Polarisation curves}
\begin{figure}
\psfig{file=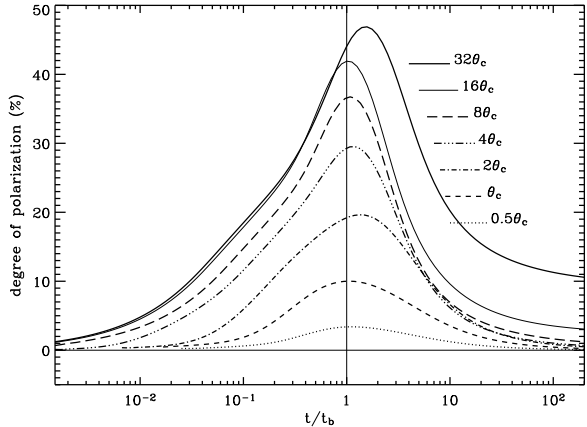,width=0.48\textwidth}
\caption{{ Polarisation curves corresponding to the lightcurves given
in Fig.~\ref{fig:nseinlc}. The x-axis for each curve is
$t/(t_{b}(\frac{\theta_o}{\theta_c}))$, where $t_b$ is found by
modelling the corresponding lightcurve with a SBP.  The break times
are given in Tab.~\ref{tab2} for $\theta_o>\theta_c$. For
$\theta_o\le\theta_c$ we use: $t_b(0)=0.026$ days, $t_b(0.5)=0.08$
days and $t_b(1)=1.12$ days; these latter break times mark only the
beginning of the second power-law branch, missing the pre-break 
slope. At $\theta_o= 32\,\theta_c$ 
the flattening in the lightcurve makes the
measurement of the break time uncertain by more than a factor of
2 (see also discussion on the measurement of $t_b$ in \S 3.2)
thus, despite the impression given
by the figure, the time of the polarisation peak for
$\theta_o/\theta_c=32$ is consistent with $t_b(32)$.}
\label{fig:nseinpo} }
\end{figure}

\begin{figure} 
\psfig{file=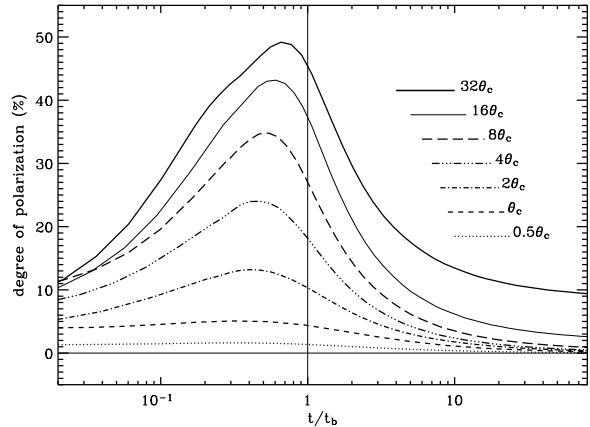,width=0.48\textwidth}
\caption{{The polarisation curves for a structured 
jet with parameters given in Fig.~\ref{fig:nseinlc} that undertakes a
sideway expansion given by Eq.~\ref{eq:sejay} with $R=0.1$. The break
times are given in Tab.~\ref{tab2} for $\theta_o>\theta_c$. For
$\theta_o\le\theta_c$ we use: $t_b(0)=0.06$ days, $t_b(0.5)=0.092$
days and $t_b(1)=0.3$ days; these latter break times mark only the
beginning of the second power-law branch, missing the pre-break
slope. In this case the corresponding lightcurves show a more
pronounced flattening and the determination of the break time becomes
more precise. All the times of the polarisation peaks are consistent
with the break times in the lightcurves, since they are shifted in the
figure by less than a factor of $2$ (see discussion on the measurement
of $t_b$ in \S 3.2).}
\label{fig:jaypola}}
\end{figure}

The polarisation curves of a structured jet with $\alpha_{\epsilon}$=2
and $\beta_{\Gamma}=$0 present all the same main features, regardless
the adopted comoving lateral speed.  As examples, we show
 a $NSE$ jet (Fig.~\ref{fig:nseinpo}) and a
jet expanding with a supersonic lateral velocity (Eq.~\ref{eq:sejay};
Fig.~\ref{fig:jaypola}). The $SE$ jets exhibit a lower degree of
polarisation and wider peaks than $NSE$ jets; in particular for a
supersonic lateral velocity, the lateral expansion starts earlier than
$\theta_{jet}(r)\sim 1/\Gamma$ (see Fig.~\ref{fig:tetjr}) and a higher
degree of polarisation is present at very early times.  Three major
features characterize all the polarisation curves resulting
from such SJs:

\begin{itemize}

\item There is only {\it one maximum} in the polarisation curve. 
Since the jet is intrinsically inhomogeneous the degree of
polarisation is greater than zero, albeit small, even at early
times. As the visible area ($\sim 1/\Gamma$) increases, the brighter
inner part weights more in the computation of the total polarisation
(Eqs.~\ref{eq:Q},~\ref{eq:U} and ~\ref{eq:P}) that increases until
$1/\Gamma\sim \theta_o$.  This is the configuration with the largest
degree of asymmetry within the visible area and therefore it coincides
with a maximum in the polarisation curve. As $1/\Gamma$ becomes larger
than $\theta_o$ the degree of asymmetry decreases along with the
polarised flux.

\item The polarisation angle 
{\it does not change} throughout the afterglow phase.  Since the
brightest spot is always at the same angle from the line of sight, the
polarisation angle does not change through the evolution of the jet.

\item The {\it maximum of polarisation decreases with $E_{iso}$}.
The larger $\theta_o$, the larger is the visible area when
$\theta_o=1/\Gamma$ and the observer sees simultaneously the bright
spine and the very dim wings so that the asymmetry is bigger.

\end{itemize}

\subsubsection{Multiwaveband polarisation}

\begin{figure}
\psfig{file=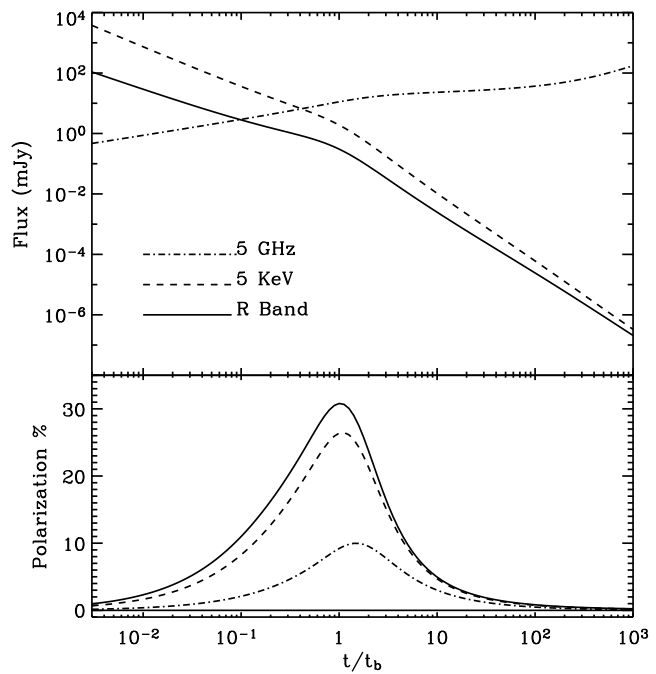,width=0.48\textwidth}
\caption{{Lightcurves (upper panel) and polarisation curves (lower panel) 
for a structured NSE jet seen at $\theta_{o}=3\theta_{c}$ in three
different frequency ranges: $\nu_a<\nu_o<\nu_m$ (dot-dashed line)
$\nu_m<\nu_o<\nu_c$ (solid line) and $\nu_c<\nu_o$ (dashed line).
All the other parameters are given in Fig.~\ref{fig:nseinlc}.
$t_{b}\simeq 1$ day.}
\label{fig:inxr}}
\end{figure}

In Fig.~\ref{fig:inxr} we compare polarisation curves (lower panel)
for the same jet configuration (the lightcurves are shown in the upper
panel) observed at $\theta_o=3\,\theta_c$ in different spectral
ranges: $\nu_a<\nu_o<\nu_m$, $\nu_m<\nu_o<\nu_c$ and $\nu_c<\nu_o$. In
the following we refer to them as defined in \S 3.3.4. The
polarisation in the radio branch is significantly smaller than in the
optical and $X$-ray bands for most of the time.  As for the HJ (see
discussion in \S 3.3.4), the peak frequency will eventually cross the
radio band and the polarisation will increase and shift on top of the
curve corresponding to $\nu_m<\nu_o<\nu_c$, while the lightcurve will
decrease following the optical lightcurve slope.  Different from the
HJ, the optical flux has an higher degree of polarisation than the
$X$-ray flux for most of the jet evolution.  We conclude that
polarisation curves depend on the spectrum and in particular in the
radio band where two peaks are generally present and the degree of
polarisation is significantly smaller than in higher frequencies bands.

\begin{table}
\centerline{\begin{tabular}{|c|cc|cc|cc|cc|cc|cc|}
\hline
& \multicolumn{2}{c}{NSE}&  \multicolumn{2}{c}{SE Eq.~\ref{se1}} & \multicolumn{2}{c} {SE Eq.~\ref{eq:sejay}} \\ 
\hline
$\theta_o/\theta_c$ & $\frac{t_b}{t_b(2)} $ & $s$ &  $\frac{t_b}{t_b(2)} $ & $s$ & $\frac{t_b}{t_b(2)}$&$s$ \\
2 & 1        & 1.29  & 1      & 1.01 &  1        & 2.63         \\
4 & 5.53     & 3.72  & 1.72   & 0.68 &  4.50    & 4.04    \\
8 & 31.2     & 12.19 & 7.66   & 1.71 & 22.30& 12.91    \\
16 & 187.53 &  20    & 37.22  & 20   & 119.50    & 20          \\
32 & 846.33 & 20     & 171.74 & 20   &  699.06   & 20    \\
64 & 4602.53  &20    & 739.32 & 20   & 3762.45   & 20 \\
\hline
\end{tabular}} 
\caption{This table summarises some results from 
the modelling with a SBP (Eq.~\ref{eq:fit}) of the lightcurves shown
in Fig.~\ref{fig:nseinlc} (NSE jet) and Fig.~\ref{fig:seinlc} ($SE$
jet) for a structured jet. The values are given for $\theta_o>\theta_c$
since for smaller viewing angles the lightcurve can only be fitted
very roughly by a SBP, missing the pre-break branch.  The break time
$t_b$ increases with the viewing angle and the general trend of all
the measured break times can be fitted by a power-law $t_b \propto
(\theta_o/\theta_c)^2$.  The smoothness parameter $s$ increases and
then saturates around $\theta_o/\theta_c=8$. A NSE jet has sharper
jet breaks compared to a SE jet.}
\label{tab2}
\end{table}

\begin{figure}
\psfig{file=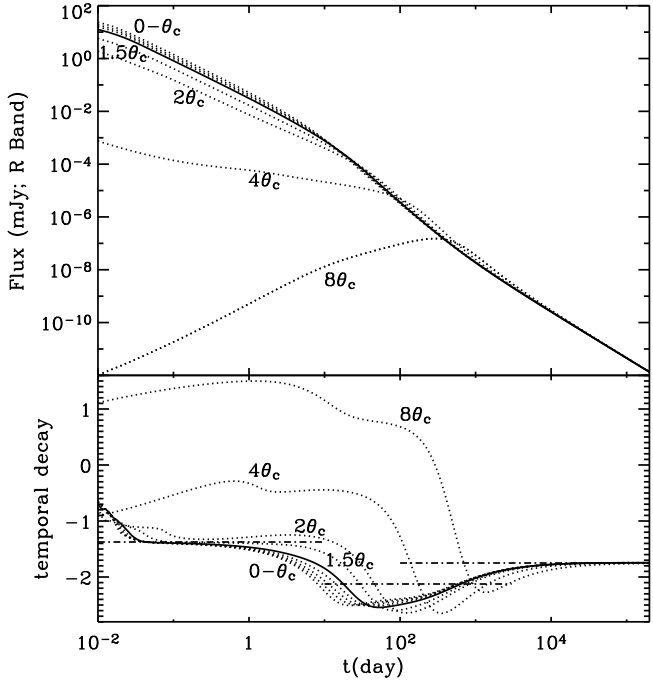,width=0.48\textwidth}
\caption{ Lightcurves in the $R$ band and temporal index for a 
Gaussian NSE jet with parameters $\epsilon_0=10^{53}$~erg,
$\Gamma_0=10^4$, $\theta_c=10\degr$, $\theta_{jet}=90\degr$,
$\epsilon_e=0.01$, $\epsilon_b=0.005$ and $n=10$~cm$^{-3}$. The
viewing angle $\theta_o$ is indicated in the figure in units of
$\theta_c$. In the upper panel, lightcurves with $\theta_o\le\theta_c$
are indistinguishable and are therefore labelled all together. The
solid curve has $\theta_o=\theta_c$. Horizontal dashed lines in the
lower panel indicate the asymptotic slopes in the various regimes.
\label{fig:gau_lc}}
\end{figure}

\begin{figure}
\psfig{file=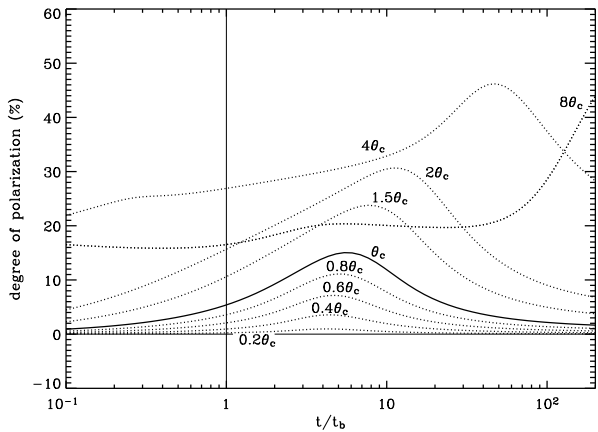,width=0.48\textwidth}
\caption{ Polarisation curves corresponding to the lightcurves shown 
in Fig.~\ref{fig:gau_lc}. The x-axis is plotted in units of $t_b(0)$,
the break time of the on-axis lightcurve, as in Figs.~\ref{fig:nse_po}
and~\ref{fig:se_po}.
\label{fig:gau_po}}
\end{figure}

\section{Gaussian jet}

 We end our investigation of GRB jet luminosity structures with a
brief discussion of the Gaussian jet curves, which present features of
both the HJ and the SJ (see also Salmonson et al. 2003).

\subsection{Presentation and Dynamics}

Perhaps a more realistic version of the standard ``top hat'' model is
a jet whose emission does not drop sharply to zero outside the
characteristic angular size ($\theta_{c}$). Such a configuration can
be described with a Gaussian distribution of the energy per unit solid
angle (Eq.~\ref{eq:gaus}).  In this jet the luminosity varies slowly
within $\theta_{c}$ ($\frac{\epsilon(0)}{\epsilon(\theta_c)}\simeq
1.65$) and it decreases exponentially for $\theta>\theta_{c}$.  In
principle, the dynamics of a Gaussian jet could be non-trivial due to
the fact that, unlike the universal SJ ($\alpha_{\epsilon}=2$) and, of
course, the HJ matter at different angles does not start spreading at
the same radius (in particularly for $\theta>\theta_{c}$); this can
develop lateral velocity gradients and transversal shock waves.  For
this reason we restrict ourselves to the case of a non-spreading
jet. This assumption is supported by recent numerical hydrodynamical
simulation for a Gaussian jet evolution; they suggest that the
transverse velocity remains below the sound speed as long as the
evolution is relativistic (Granot \& Kumar 2003).

\begin{table}
\centerline{\begin{tabular}{|c|cc|cc|}
\hline
& \multicolumn{1}{c}{Gaussian NSE} \\ \hline
$\theta_o/\theta_{c}$ & $\frac{t_b}{t_b(0)} $ & $s$  \\ \hline
0 & 1 & 2.46 \\
0.2 & 1.03 &  2.38\\
0.4 & 1.14 &  2.28 \\
0.6 & 1.34 &  2.16\\
0.8 & 1.66 &  2.25 \\
1.0 & 2.13 & 2.55 \\
1.5 & 3.82 &  4.36 \\
2.0 & 5.87 &  5.73 \\
4.0 & 19.04 & 3.16 \\
8.0 & 72.39 & 0.91 \\
\hline
\end{tabular}}
\caption{ The same as Tab.~1 for the lightcurves shown in
Fig.~\ref{fig:gau_lc} ($NSE$ jet).  The break time $t_b$ is given as a
multiple of the $t_b$ fitted by an on-axis observer [$t_b(0)=4.56$
days]. }
\label{tabgau}
\end{table}

\subsection{Lightcurves}
 The resulting lightcurves are shown in Fig.~\ref{fig:gau_lc}.  It
 shows that for $\theta\le\theta_{c}$ the lightcurves are very similar
 to a HJ's ones: the slopes, the values and the behaviour of the break
 time as a function of angle (see Tab.~\ref{tabgau}) are consistent
 with what expected for a NSE HJ seen within the cone (see
 Tab.~\ref{tab}). On the contrary the break does not become smoother
 (3rd column of Tab.~\ref{tabgau}) as the line of sight approaches
 the edge, like in the HJ but its shape remains rather unaltered.
 Finally, on average the Gaussian jet has a less sharp break in the
 lightcurve.  This is in agreement with previous calculations (Granot
\& Kumar 2003).  For $\theta>2\theta_c$ instead the pre-break slope
becomes flatter and flatter and for $\theta\ge 8\theta_c$ it becomes
positive; in this latter case the lightcurve is actually dominated by
the emission coming from the core.
  
\subsection{Polarisation curves}
The polarisation curves present intermediate characteristics between the
HJ and the SJ ones (Fig~\ref{fig:gau_po}).  The absence of edges and
the presence of a symmetric luminosity gradient with respect to the
jet axis, produce (as in the case of a SJ) a one-peak curve with a
constant polarisation angle.  On the other hand, the exponential
decrease of luminosity outside the core makes the relation between
polarisation curve and lightcurve more close to the HJ's one: the peak
is located after the break time in the total flux.  As the viewing
angle increases, especially for $\theta_o>\theta_{c}$ the maximum in
the lightcurve moves torwards $t=t_b$, but eventually the core starts
to dominate the lightcurve since early times because of the
exponential luminosity distribution and the polarisation curves
resembles that of an orphan afterglow.

\section{Comparison and Discussion}
The previous sections confirm what first claimed in RLR02: the
lightcurves of a HJ and of SJ with $\alpha_{\epsilon}=2$ are very
similar but their intrinsic features depend on the viewing angle for
the SJ and on the opening angle for the HJ. On the other hand the
polarisation curves are completely different.  These characteristics
are the direct consequences of an energy distribution
$\epsilon\propto\theta^{-2}$: the lightcurve is dominated by the line
of sight emission while (for any $\alpha_{\epsilon}$) the polarisation
curve is dominated by the emission coming from an angle $1/\Gamma$
with the line of sight. This means that the total flux we receive does
not bear footprints of the jet structure while the observed
polarisation does. In Fig.~\ref{fig:inomcom} we directly compare the
lightcurves and the polarisation curves from HJ and SJ with
$E_{iso}(\theta_o)=E_{iso}$ and $\theta_{jet}=2\,\theta_o$.  These are
the parameters for which their lightcurves are more similar.  We
also show the lightcurve and polarisation curve for a Gaussian jet
with $E_{iso}(\theta_o)=E_{iso}$ and
$\theta_{c}\simeq0.6\,\theta_{jet}$.

\subsection{Lightcurves}
 Fig.~\ref{fig:inomcom} (upper panel) summarises the comparison
discussed \S 4.1 and \S 5.2 among the characteristics of the
lightcurves of the three jet structures.  In particular it should be
noticed the flattening before the break, present in the SJ lightcurve
and the almost perfect match between the HJ and GJ lightcurves (the GJ
lightcurve has been divided by a factor 2 or it would be overlaid don
the HJ one). The pre-break bump is actually the only sign of the
underlying jet structure: when $1/\Gamma\sim\theta_o$ and the
lightcurve breaks, the jet core is also visible and its contribution
gives that flux excess that we perceive as a flattening. The larger is
the viewing angle the smaller is $E_{iso}$ and more prominently the
core out-shines the line of sight flux at the break. Some authors
claimed (e.g. Granot \& Kumar 2003) that this feature can used to
discriminate between the HJ and the SJ. However the shape and the
intensity of the flattening depends on how the wings and the core join
together and this is a free parameter
($\beta_{\epsilon}$). Consequently comparing the model with
observations provides a way to fix the shape of the energy
distribution but not a way to test the model itself.  Besides the
pre-break flattening the temporal behaviour of the lightcurves plotted
in Fig.~\ref{fig:inomcom} is identical and for this reason the same
data can be fitted with any of the three models (e.g Panaitescu
\& Kumar 2003, for SJs and HJs).

\subsection{Polarisation curves}
\begin{figure}
\psfig{file=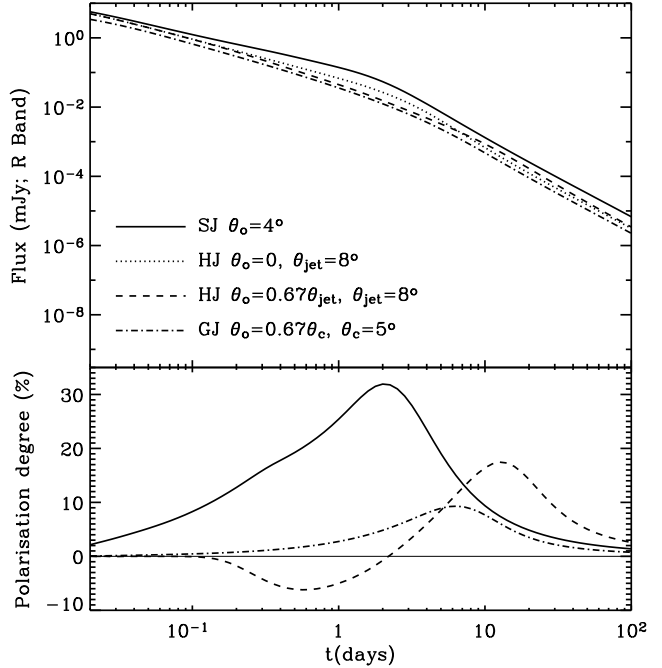,width=0.48\textwidth}
\caption{{Lightcurves (upper panel) 
and polarisation curves (lower panel)
comparison between a SJ seen at $\theta_o=4^{\circ}$
(the other parameters are given in Fig.~\ref{fig:nseinlc}),
 a homogeneous jet with $\theta_{jet}\simeq
2\,\theta_{o}$, $E_{iso}\simeq E_{iso}(\theta_o)$ and 
$\Gamma_0\simeq\Gamma_0(\theta_{o})$ and a
Gaussian jet with the same $E_{iso}$ and $\Gamma_0$ but characteristic
size $\theta_c\simeq0.6\theta_{jet}$ ; the HJ is seen on-axis and at 
$\theta_o=0.67$. Both jets do not undertake lateral expansion.}
\label{fig:inomcom}}
\end{figure}

The comparison in Fig.~\ref{fig:inomcom} (lower panel) clearly shows
the main differences between the polarisation curves of a SJ,  of
a GJ and of a HJ:

\begin{itemize}
\item {\it Different behaviour from the beginning.}

\noindent
The HJ  and GJ do not display polarisation at early times but
only when the visible area intersects the edge of the jet and an
asymmetry is present.  However, the SJ shows, from the beginning,
regions with different luminosity within $1/\Gamma$ from the line of
sight and therefore we observe a non zero, albeit small, degree of
polarisation even at very early times.

\item {\it Different evolution of the polarisation angle: $90^{\circ}$
rotation vs constant}

\noindent
The polarisation angle for an HJ rotates by $90^{\circ}$ at $\sim
t_{b}$; this is because opposite parts of the jet dominate the total
polarisation when $1/\Gamma$ overtakes the nearer edge and when it
reaches the furtherest one.  For a jet seen off-axis the break in the
lightcurve is not sharp (\S 3.1.1) and the time break $t_{b}$ occurs
roughly when $\Gamma\sim(\theta_o+\theta_{jet})/2$, midway between the
two edges.  That is why the change in the polarisation angle
corresponds to the jet break.  This behaviour is observed for any
off-axis angle, for all sideways expansions considered in this
paper. On the other hand in the SJ  and GJ the same spot (towards
the core) always dominates the total polarisation and consequently the
angle remains constant.
  
\item {\it Different number of peaks: 2 vs 1}

\noindent
The HJ has two peaks in the polarisation lightcurve, with the second
always higher than the first.  This is of course related to the
presence of the two edges of the jet that are the source of the
asymmetry within the visible area (Fig. 1 and 2 GL99). The SJ has only
one maximum when the observer sees light coming from the core at
$1/\Gamma$ from the line of sight. This is when a break occurs in the
lightcurve and the polarised flux is almost completely dominated by
the flux coming from the more powerful region of the jet.  The
behaviour of the GJ is in between: the polarisation angle is constant
as for a SJ, but the polarisation peak is delayed with respect to the
break time, as in a HJ.

\end{itemize}

\section{Conclusions}

We have presented a thorough analysis of light and polarisation curves
from GRB jets. Our study is based on codes that numerically
integrate the jet equations taking into account the equal arrival
time surfaces and the finite width of the radiating shell. We
initially concentrate on homogeneous jets, pointing out some features
of the jet evolution which had been overlooked in previous works.  In
particular we underline the difficulty of measuring the break time and
the electron energy distribution index $p$ modelling the lightcurve
with a SBP; we also point out that in the non-relativistic regime the
usually assumed Sedov-Taylor solution could be not correct when
applied to narrow jets. Concerning polarisation, we find the presence
of only two peaks in the polarisation curve, for any off-axis angle
and for any sideway expansion velocity considered so far in
literature.

 The main result of this paper is that the polarisation temporal
behaviour is found to be very sensitive to the luminosity distribution
of the outflow unlike the total flux curve.  We have achieved the goal
of calculating, for the first time, the polarisation from a universal
structure jet and from a Gaussian jet and performing a comparison (of
both lightcurves and polarisation curves) with what is predicted by
the standard jet model.

 In particular the derivation of the light and polarisation
curves for structured jets has been performed under several assumption
for the dynamics. For simplicity we study only a non-spreading
GJ.  Our numerical lightcurves confirm the previous conclusion
(RLR02; Granot \& Kumar 2003; Kumar \& Granot 2003; Salmonson 2003)
that, based on the lightcurve properties, it is extremely hard to
infer the structure of the jet.  Polarisation curves, on the other
hand, are extremely different. Since the brightest part of the jet is
always on the same side for structured (SJ and GJ) outflows, the
position angle of polarisation remains constant throughout the whole
evolution. In addition, for a SJ the polarisation peaks coincident in
time with the jet break in the light curve, which instead corresponds
to the time of minimum polarisation in homogeneous jets. The
exponential wings in a GJ, however, shift the position of the peak
after the break in the lightcurve, a feature that marks the difference
between the GJ and the SJ predicted polarisation. We should however
stress that due to the many uncertainties inherent in the derivation
of polarisation curves, it is hardly possible to use them to measure
in a fine way the energy distribution of the jet (e.g. tell a
$\epsilon\propto\theta^{-2}$ structured jet from a
$\epsilon\propto\theta^{-2.5}$ one). What polarisation can robustly
determine is whether the energy distribution in the jet is uniform or
centrally concentrated. Alternative approaches, such as the observed
luminosity function (RLR02) are also important to further constrain
the jet profile, even though the data seem not to be accurate enough
at this stage (Perna et al. 2003; Nakar et al. 2004).
 
Finally we present unprecedented polarisation curves
corresponding to different spectral branches, both for a HJ and for a
SJ: we find a spectral dependence of the degree of polarisation and we
expect changes in the temporal behaviour of the polarised fluxes (in
all bands but in particular in the radio band) associated with
spectral breaks in the lightcurve.

It should be emphasised that these results hold in the absence of
inhomogeneities in the external medium and/or in the luminosity
distribution within the jet (other than $\epsilon \propto
\theta^{\alpha_{\epsilon}}$).  Any breaking in the fireball symmetry
causes random fluctuations of both the value and the angle of the
polarisation vector (Granot \& K\"onigl 2003; Lazzati et al. 2003;
Nakar \& Oren 2004).  In the lightcurve the main consequence is the
presence of bumps and wiggles on top of the regular power-law decay
(e.g. Lazzati et al. 2002; Nakar, Piran \& Granot 2003).  Actually a
complex behaviour for the lightcurve has always been observed so far
together with a peculiar polarisation curve, such as in GRB 021004
(Rol et al. 2003, Lazzati et al. 2003) and in GRB 030329 (Greiner et al.
2003). On the other hand GRB 020813 presents an extremely smooth
lightcurve (Gorosabel et al. 2004; Laursen \& Stanek 2003) associated
with a very well sampled polarisation curve (Gorosabel et al. 2004)
that is characterised by a constant position angle and a smoothly
decreasing degree of polarisation.  This burst is thus particularly
suited for a proper comparison with data of the models described in
this paper.  Lazzati et al. (2004) has performed the modelling of the
polarisation curve according to several models (including all the
models considered here) and they find that the structured model can
successfully reproduce the data and it can predict the jet-break time
in agreement with what measured in the lightcurve.

Further complication can arise from the presence of a second non
negligible coherent component of the magnetic field in the ISM. Our
results have been obtained assuming that the magnetic field
responsible for the observed synchrotron emission is the one generated
at the shock (thus tangled at small scales).  This is a reasonable
assumption since the compression of a standard interstellar field is
far too low to produce the observed radiation. However if the burst
explodes in a pre-magnetised environment or if the jet is magnetic
dominated and the field advected from the source survives till the
afterglow phase, then the polarisation curve will be the result of the
relative strength of the two components of the magnetic field.  Granot
\& K\"onigl (2003) have discussed the polarisation curve in the former
case for an homogeneous jet propagating through a magnetic wind
bubble.

Finally the intrinsic polarisation curve of the afterglow can be
affected by the dust present both in the Milky Way and in the host
galaxy. Lazzati et al. 2003 discuss the modification of the transmitted
polarised vector and they show that in GRB 021004 a sizable fraction
of the observed polarised flux is likely due to Galactic selective
extinction.

Despite the difficulties inherent to polarisation observations and
modelling (Lazzati et al. 2004), we believe that polarimetric studies
are of great importance in determining the structure of GRB outflows.

\section*{Acknowledgements}
We thank Martin J. Rees for useful and stimulating discussions and
Jonatan Granot for numerous interaction and comparisons between our
results. ER thanks the Isaac Newton and PPARC studentships for
financial support. DL acknowledges support from the PPARC postdoctoral
fellowship PPA/P/S/2001/00268.

\end{document}